\documentclass{elsarticle}[12pt]
\usepackage{epsf}
\usepackage[latin1]{inputenc}
\usepackage{amsmath,amsfonts,amssymb}

\date{}

\begin{document}

\begin{titlepage}
\begin{center}
{\large\bf
Clustering properties, Jack polynomials  and unitary conformal field theories. 
}\\[.3in] 

{\bf Benoit Estienne$^{1}$, Nicolas Regnault$^{2}$, Raoul\ Santachiara$^{3}$}\\
	$^1$ {\it LPTHE, CNRS, UPMC Univ Paris 06 \\
Bo\^ite 126, 4 place Jussieu, F-75252 Paris Cedex 05\\
            e-mail: {\tt estienne@lpthe.jussieu.fr}}\\

    $^2$ {\it Laboratoire Pierre Aigrain, Departement de Physique, ENS, CNRS,\\
24 rue Lhomond, 75005 Paris,  France\\
    e-mail: {\tt regnault@lpa.ens.fr}}\\

	$^3$ CNRS, {\it LPTMS,
             Universit\'e Paris-Sud,\\UMR 8626, 
             B\^atiment 100\\
             91405 Orsay, France. \\
    e-mail: {\tt raoul.santachiara@lptms.u-psud.fr}. }\\
\end{center}
%\vskip .04in
\centerline{(Dated: \today)}
\vskip .2in
\centerline{\bf ABSTRACT}
\begin{quotation}
Recently, Jack polynomials have been proposed as natural generalizations of $\mathbb{Z}_k$ Read-Rezayi states describing non-Abelian fractional quantum Hall systems.  These polynomials are conjectured to be related to  correlation functions of a class of W-conformal field theories based on the Lie algebra $A_{k-1}$.   
These theories can be considered as  non-unitary solutions of a more general series of CFTs with $ \mathbb{Z}_k $ symmetry, the parafermionic theories. Starting from the observation that some parafermionic theories admit unitary solutions as well,  we show,  by computing the corresponding correlation functions,  that these theories  provide trial wavefunctions which satisfy the same clustering properties as the non-unitary ones.
We show explicitly that, although the  wavefunctions constructed by unitary CFTs cannot be expressed as a single Jack polynomial, they still show a fine structure where the mathematical properties of the Jack polynomials  play a major role.
%\begin{quotation}
\end{quotation}
\end{titlepage}

\vskip 0.5cm
\noindent
%\pacs

\section{Introduction}
Since the success of the Laughlin states \cite{Laughlin}, the use of trial wavefunctions in the fractional quantum Hall (FQH) effect has provided deep insights into these systems, especially non-Abelian ones \cite{MooreRead,ReadRezayi}.  Over the past few years there has been a renewed interest in non-Abelian states mostly because of their connection with topological quantum computing \cite{review_topo}.  Model wavefunctions describing non-Abelian states can be constructed by using the conformal blocks of conformal field theories (CFTs).  Much of the theory  underlying the non-Abelian states is based on the monodromy properties of the conformal blocks. 

The Read-Rezayi (RR) states\cite{MooreRead,ReadRezayi}, some of which are thought to be experimentally relevant, are a paradigm for non-Abelian states. These states are given by the conformal blocks of a particular family of CFTs, the so-called $\mathbb{Z}_k$ Fateev-Zamolodchikov (FZ) parafermionic field theories (defined below) \cite{FZ}.  These are CFTs with an extended cyclic $\mathbb{Z}_k$ symmetry to which corresponds  a set of conserved current with  a particular operator product expansion (OPE).   The current OPEs define the so-called  parafermionic algebras and the wavefunctions constructed by using the corresponding current correlation functions inherit specific clustering properties. In particular one can show that, apart from some gauge factor, the (bosonic) $\mathbb{Z}_k$ RR ground states are symmetric polynomials which vanish when at least $k+1$ particles come to the same point. The $k-$clustering properties make the $\mathbb{Z}_k$ RR  states to be the unique densest zero-energy ground states of a $ k+1$ body interaction Hamiltonians describing the energy cost to have $k+1$ particle occupying the same position \cite{ReadRezayi,SRC_projham}. In the case  of the $2$-body ($k=1$) interaction, the zero-energy ground state is the Laughlin state for bosons at filling fraction $\nu=1/2$. The model Hamiltonian of this kind are believed to capture the physics of two-dimensional systems in very high magnetic fields where the effective hamiltonian is reduced to the  interaction between particles in acting in the lowest Landau level. 

Because of the importance of the RR wavefunctions, an intense research activity has been focused on the generalizations of these functions and therefore of the FZ parafermionic theories.
In addition to  the $k$-clustering properties,  symmetric polynomial are also characterized by the power $r$ with which the polynomials vanish when the $k + 1$st particle arrives. In terms of CFT, as we will see, the value of $r$ determines the conformal dimension of the currents generating the $\mathbb{Z}_k$ symmetry. The RR states have $r=2$. 
 As it was observed in \cite{BernevigHaldane1}, the $\mathbb{Z}_k$ Read-Rezayi wavefunctions, as well as other previously proposed non-Abelian wavefunctions \cite{Gaffnian}, can be written in terms of a single Jack polynomial (Jacks, defined below) with negative parameter $\alpha=-(k+1)/(r-1)$. This has naturally suggested the possibility of describing quantum Hall wavefunctions in terms of Jacks.  This approach has been the subject of a series of recent works \cite{BernevigHaldane2,BernevigHaldane3,Ardonne,BernevigW} where the connection between Jacks and FQHE models has been studied in detail.

Interestingly,  it was conjectured that the Jacks are directly related to correlators of certain CFTs based on the Lie algebra $A_{k-1}$,  the so called $WA_{k-1}(k+1,k+r)$ theories \cite{Wtheory,FJMM,BernevigHaldane2}. 

%This was known to be true for $k=2$ and arbitrary $r$ where the   $WA_{1}(3,2+r)$ coincide with  the series of minimal model $M(3,2+r)$ [..] and for $r=2$ and arbitrary $k$ as $WA_{k-1}(k+1,k+2)$ coincides with the $\mathbb{Z}_k$ FZ theory. Further evidences of the relation between Jacks and $WA_{k-1}-$ correlators for arbitrary  $k$  has been provided in [..]. 

A crucial point is that for $r>2$, the $WA_{k-1}(k+1,k+r)$ theories are non unitary as it is manifest from the negative value of their central charge $c$ .  However,   there are solid arguments \cite{Readnu1,Readnu2} that the wavefunctions constructed using non-unitary CFT cannot describe topological gapped quantum phases.  In this respect,  a recent work \cite{Thierry} has proposed that  unitary Abelian theories may be built from non-unitary ones.

The initial observation that motivates the present work is that there is a family of CFTs, the parafermionic theories $\mathbb{Z}_{k}^{(r)}$ defined below, which include the $WA_{k-1}(k+1,k+r)$ theories as a special case. 
In particular, there exist  $\mathbb{Z}_{k}^{(r)}$ theories based on a current algebra which is associative for each value of the central charge $c$, the $\mathbb{Z}_{k}^{(r)}(c)$ algebras. Analogously to the unitary sequence of minimal models based on the Virasoro algebra, the $\mathbb{Z}_{k}^{(r)}(c)$ algebras admit unitary representations for some discrete series of $c$ values. We show that the correlators of $\mathbb{Z}_{k}^{(r)}(c)$ provide trial wavefunctions which satisfy for arbitrary $c$ the same clustering properties as the non-unitary ones. Moreover, we show explicitly that, although the unitary wavefunctions cannot be expressed as a single Jack, they still show a fine structure where the mathematical properties of the Jacks play a major role.

\section{Clustering Polynomials}
\label{clustering_poly}

The CFT results presented in this paper have been inspired and motivated by the study of non-Abelian wavefunctions in FQH systems. In this section, we review some general properties of quantum Hall ground state coming from simple physical arguments.

It is convenient to place the systems of $n$ particles on a spherical surface of radius $R$ with a uniform radial magnetic field with total flux $N_{\phi}$\cite{Haldane_sphere}. 
The position of $i$-th particle on the sphere can be represented as a complex variable $z_i$ which is its stereographic projection. Each particle in the lowest Landau level has orbital angular momentum $N_{\phi}/2$ and the single-particle basis states have the form $z^m \mu(z,\bar{z})$ where $m$ is the $L_z$ momentum quantum number and $\mu(z,\bar{z})$ is the measure on the sphere, $\mu(z,\bar{z})=1/(1+(|z|/R)^2/4)^{1+N_{\phi}/2}$. Therefore many-body wavefunction $\tilde{\Psi}$\footnote{we use $\tilde{\Psi}$ to distinguish  the many-particles wavefunction from the parafermionic field $\Psi$ defined later} describing $n$ particle states in the lowest Landau levels take  the form:
\begin{equation}
\tilde{\Psi}(z_1,\bar{z}_1,\cdots,z_n, \bar{z}_n)=P_{n}(z_1,\cdots,z_n)\prod_{i=1}^{n}\mu(z_i,\bar{z}_i),
\end{equation}
where $P_{n}(\{z_i\})$ is a polynomial in the $n$ variables $z_i$. In this article, we will focus on bosonic systems (fermionic wavefunctions are related to the bosonic ones by a Jastrow factor). Therefore the polynomial  $P_{n}(\{z_i\})$ is symmetric in the variables $z_i$.

To describe a quantum Hall ground state, the polynomial $P_{n}(\{z_i\})$ has to be a translationally and rotationally invariant homogeneous polynomial  with $N_{\phi}$ being the highest power in each variable $z_i$.  This means the polynomial have to satisfy the conditions: 
\begin{eqnarray}
\sum_{i=1}^{n}\partial_i P_{n}(\{z_i\}) & = & 0 \label{HW}\\
 \sum_{i=1}^{n} z_i^2 \partial_i P_{n}(\{z_i\})& = & - N_{\phi}\left( \sum_i z_i\right)  P_{n}(\{z_i\}) \label{LW}\\
\sum_{i=1}^{n} z_i \partial_i P_{n}(\{z_i\})& = & \frac{n N_{\phi}}{2} P_{n}(\{z_i\})\label{degree}
\end{eqnarray}
where the conditions (\ref{HW})-(\ref{LW}) are usually referred respectively as the highest weight (HW) and lowest weight (LW) conditions. On the sphere geometry, $N_\phi$ and $n$ are related by the linear identity $N_\phi=\nu^{-1}n-\delta$ where $\nu$ is the filling factor and $\delta$ the so-called shift.

Here we are interested in polynomials  $P_{n}(\{z_i\})$ which are the zero-energy eigenstates of model $k+1$-body Hamiltonian. Using the notation of \cite{SRC_projham}, in the spherical geometry these Hamiltonians can be defined as the projection operator $\mathcal{P}^{r}_{k}$ that kills the state where any cluster of $k+1$ particles has relative  angular momentum less than $r$. In other word no more than $k$ particles can occupy $r$ consecutive orbitals. Defining $P_{n}^{(k,r)}(\{z_i\})$ the zero energy eigenstates of $\mathcal{P}^{r}_{k}$, one can show that the $P_{n}^{(k,r)}(\{z_i\})$ should satisfy the $k-$ clustering conditions:
\begin{eqnarray}
P_{n}^{(k,r)}(z_i=z_2=..=z_{k+1},z_{k+2},\cdots, z_{n})&=&0 \label{kclustering}\\
P_{n}^{(k,r)}(\underbrace{z_i=z_2..=z_{k}}_{\text = Z},z_{k+1},\cdots, z_{n})&=&\prod_{i=k+1}^{n}(Z-z_i)^rP_{n-k}^{(k,r)}(z_{k+1},z_{k+2},\cdots, z_{n})\label{rpower}
\end{eqnarray}

As pointed out in \cite{SRC_projham,SRC_projham2}, the highest density zero energy states satisfying the above properties can be found at
\begin{equation}
N_{\phi}=\frac{r(n-k)}{k}. 
\end{equation}
These polynomials describe (bosonic) quantum Hall states at filling fraction $\nu=k/r$, with a shift $\delta=r$.

As we will show in detail below, the polynomials $P^{(k,r)}_{n}$ satisfying the conditions (\ref{HW})-(\ref{rpower}) can be constructed by using general CFTs with extended $\mathbb{Z}_k$ symmetry. Moreover, the symmetric polynomials $P^{(k,r)}_{n}$ are naturally associated to Jacks with negative fractional parameter.
We will use these two approaches  to describe the most general properties of the polynomials  $P^{(k,r)}_{n}$.

 \section{Symmetric polynomials and Jacks at $\alpha=-(k+1)/(r-1)$}
\label{Jacks}
A general characterization of symmetric polynomials with $k-$clustering properties (\ref{kclustering}) was initiated in the work of Feigin et al. \cite{FJMM}.  In this section we briefly review their results and fix our notations.  

The Jacks $J^{\alpha}_{\lambda}(z_1,\cdots,z_n)$ are symmetric polynomials of $n$ variables depending rationally on a parameter $\alpha$ and indexed by a partition $\lambda$, $\lambda=[\lambda_1,\lambda_2\dots \lambda_n]$ where the $\lambda_i$ are positive integers and $\lambda_i \geq \lambda_{i+1}$. For more details on Jacks see \cite{Macdonald}. Defining the monomial functions $m_\lambda$ as :
\begin{equation}
m_{\lambda}(\{z_i \})=\mathcal{S} (\prod_i^n z_i^{\lambda_i}) 
\end{equation}
where the $\mathcal{S}$ stands for the symmetrization over the $n$ variables, the expansion of a Jack over the $m_{\lambda}$ basis takes the form \cite{Macdonald}:
\begin{equation}
J^{\alpha}_{\lambda}=m_{\lambda}+\sum_{\mu< \lambda} u_{\lambda \mu}(\alpha)m_{\mu}.  
\label{expan_jack}
\end{equation}
The dominance ordering $\mu\leq \lambda$ in the sum is defined as $\mu_1+\cdots+\mu_i\leq \lambda_1+\cdots\lambda_i$ ($1\leq i\leq n)$.
The Jacks are eigenfunctions of a Calogero-Sutherland Hamiltonian \cite{Sutherland}:
\begin{equation}
\left[\sum_{i}^{n}(z_i\partial_i)^2+\frac{1}{\alpha}\sum_{i<j}\frac{z_i+z_j}{z_i-z_j}(z_i\partial_i - z_j \partial_j)-\varepsilon_{\lambda}\right]J^{\alpha}_{\lambda}(z_1,\cdots,z_n)=0
\label{CS}
\end{equation}
with eigenvalues $\varepsilon_{\lambda}=\sum_{i}^n \lambda_i(\lambda_i+1/\alpha(n+1-2i))$.

Let $k,r$ be positive integer such that $k+1$ and $r-1$ are co-prime. A  partition $\lambda$ is said to be $(k,r,n)$-admissible if it satisfies the following condition:
\begin{equation}
\lambda_{i}-\lambda_{i+k}\geq r \quad (1\leq i\leq n-k).
\label{admissible_part}
\end{equation}
Given a $(k,r,n)$-admissible partition $\lambda$ Feigin {\& al}\cite{FJMM} showed that:
\begin{itemize} 
\item the coefficients $u_{\lambda \mu}(\alpha)$ do not have a pole for the particular negative value $\alpha=-(k+1)/(r-1)$.

\item the  Jack  $J^{-(k+1)/(r-1)}_{\lambda}(z_1,\cdots,z_n)$ vanishes when $z_1=z_2=\cdots=z_{k+1}$.

\item Let $I^{(k,r,n)}$ be the space spanned by the Jacks $J^{-(k+1)/(r-1)}_{\lambda}(z_1,\cdots,z_n)$ for all $(k,r,n)$-admissible partitions $\lambda$. $I^{(k,2,n)}$ coincides with the space of symmetric polynomials satisfying (\ref{kclustering}).
\end{itemize}
 The  case where the polynomials $P_{n}^{(k,r)}(\{z_i\})$ are  given by a {\it single}  Jack $J^{\alpha}_{\lambda}(\{z_i\})$ was considered  in \cite{BernevigHaldane1}. It was shown in \cite{BernevigHaldane1} that   a necessary condition for a  Jack $J^{\alpha}_{\lambda}(\{z_i\})$ of total degree $n N_{\phi}/2$ to satisfy  the Eq.(\ref{HW}) {\it and} the  Eq.(\ref{LW}) is  to have i) $\alpha=-(k+1)/(r-1)$ negative rational  and ii) $\lambda$ a $(k,r,n)$- admissible partition.   For a given a flux  $N_{\phi}=r(n/k-1)$, the only  $(k,r,n)$- admissible $\lambda$ is:
\begin{equation}
\lambda=[\underbrace{N_{\phi},N_{\phi},..}_{\text  k -\mbox{times}}, \underbrace{N_{\phi}-r,N_{\phi}-r,..}_{\text k -\mbox{times}},\cdots ,\underbrace{r,r,..}_{\text  k -\mbox{times}}]
\label{only_admissible_lambda}
\end{equation}
Finally, let  $s$ be an integer greater than one, $s>1$,  $n_{0}=s(k+1)-1$ (corresponding to the number of particles occupying the $m=0$ orbital \cite{BernevigHaldane1}) and $\lambda$ being  a $(k,r,n-n_{0})$-admissible partition, defined above.  
 If one releases the (LW) condition but keeps the (HW) one, the Jacks are still characterized by having a negative $\alpha=-(k+1)/(r-1)$ and  a partition  $\lambda'$  which is obtained from $\lambda$ by putting $\lambda_{i}^{'}=\lambda_{i}+s(r-1)+1$ for $i=1,2\cdots n-n_{0}$. These states are interpreted  as  the quasi-holes states of the $\nu=k/r$ FQH states \cite{BernevigHaldane1}.

\section{Parafermionic theories and $k-$clustering symmetric polynomials}
A powerful method\cite{MooreRead,ReadRezayi} to generate polynomials which satisfy the properties (\ref{HW})-(\ref{rpower}) is based on the chiral current correlation functions of the parafermionic theories.  These are  $\mathbb{Z}_k$ invariant CFTs where the current algebra is generated by a set of $k$ parafermionic fields $\{\Psi_{q}(z)\}_{q=0,1,\cdots,k-1}$, $\Psi_{0}=I$ being the identity operator, with conformal dimensions $\Delta_{q}=\Delta_{k-q}$. The parafermionic algebra takes the general form:
\begin{eqnarray}
\Psi_q(z) \Psi_{q'}(w)&=&\frac{\gamma_{q,q'}}{(z-w)^{\Delta_{q}+\Delta_{q'}-\Delta_{q+q'}}}[\Psi_{q+q'}(w)]
\label{para_algebra1}\\
\Psi_q(z) \Psi_{k-q}(w)&=&\frac{1}{(z-w)^{2\Delta_{q}}}\left( 1+(z-w)^2\frac{2\Delta_{q}}{c} T(w)+\cdots \right)
\label{para_algebra2}
\end{eqnarray}
where the sums $q+q'$ are defined modulo $k$, $[\Psi]$ indicates the operator $\Psi$ and its Virasoro descendants and the $\gamma_{q,q'}$ are the algebra coupling constants. 
The above algebra realizes the extended $\mathbb{Z}_k$ symmetry and includes the conformal one, as it can be seen from the presence of the stress-energy operator $T(w)$ in (\ref{para_algebra2}).  

We stress that the current fusions (\ref{para_algebra1})-(\ref{para_algebra2}) represent only the general form for a candidate parafermionic algebra. The difficult problem is to define completely the form of these OPEs, by fixing the values of $\Delta_{q}$ and all the singular terms in the current expansions in order to obtain an associative algebra. Associativity ensures that the  correlation functions of the chiral currents  $\{\Psi_j\}$ are well defined.

For a given candidate associative $\mathbb{Z}_k$ algebra, the value of the dimensions $\Delta_{q}$ of the field $\Psi_q$  is the initial input. The simplest series of $\Delta_{q}$ dimensions that satisfy the first associative constraints are :

\begin{equation}
\Delta_{q}=\frac{r}{2}\frac{ q(k-q)}{ k}  \quad q=1,\cdots,k-1; \;r=2,3\cdots
\label{dim_ass}
\end{equation}
Therefore, for a given $\mathbb{Z}_k$ symmetry, the possible set of values of the current dimensions are determined by the integer $r$. Associativity requirements lead to a great number of additional constraints. If, for a value of $r$ an associative algebra can be found,
it is easy to verify that the polynomials $P_n^{(k,r)} (\{z_i\})$  defined by :
\begin{eqnarray}
P_n^{(k,r)} (\{z_i\})& \hat{=} & \langle \Psi(z_1) \hdots  \Psi(z_{n}) \rangle \prod_{i<j} \left( z_i-z_j \right)^{2\Delta_1-\Delta_2} \nonumber \\
& = & \langle  \Psi(z_1) \hdots  \Psi(z_{n}) \rangle \prod_{i<j} \left(z_i-z_j \right)^{r/k}.
\label{CFT_FQH}
\end{eqnarray}
where $\Psi(z)\hat{=}\Psi_1(z)$, satisfy the conditions (\ref{HW})-(\ref{rpower}). In this case one assumes that the number of particles $n$ is a multiple of $k$ (otherwise the correlation function is zero).  The HW and LW conditions, see Eqs.(\ref{HW})-(\ref{LW}),  are a direct consequence of the global conformal invariance of CFT correlators while the $k-$clustering properties,  Eqs(\ref{kclustering})-(\ref{rpower}),  are given by the $\mathbb{Z}_k$ symmetry of the current algebra \cite{MooreRead}  (see Appendix \ref{AppConf}).

The study of the symmetric polynomials $P_{n}^{(k,r)}(\{z_i\})$ is then directly related to the existence of  associative solutions of the algebra (\ref{para_algebra1}),(\ref{para_algebra2}). This close connection between CFT approach and the classification of clustering polynomials has been also pointed out in \cite{Wen1,Wen2,Wen3}. These solutions will be referred as $\mathbb{Z}_k^{(r)}$ theories. We will now discuss in more details this family of models and their properties.

\section{$\mathbb{Z}_k^{(r=2)}$: Fateev-Zamolodchikov parafermionic models and Read-Rezayi states}
\label{FZt}
The number of the singular terms  in (\ref{para_algebra1})-(\ref{para_algebra2}), and thus the complexity of the algebra,  increases with  $r$. The case $r=2$ is the simplest one and it is determined by the following fusion rules for the $\Psi_1$ and $\Psi_{k-1}$ fields :
\begin{eqnarray}
\Psi_1(z) \Psi_{1}(w)&=&\frac{1}{(z-w)^{2/k}}\Psi_{2}(w)+\mbox{regular terms}
\label{para_algebra1_RR}\\
\Psi_1(z) \Psi_{k-1}(w)&=&\frac{1}{(z-w)^{2(k-1)/k}}\left(I+\frac{k+2}{k}(z-w)^2 T(w)\right)+\mbox{regular terms}
\label{para_algebra2_RR}
\end{eqnarray}
The above algebra is associative with a fixed value of the central charge $c=2(k-1)/(k+2)$. The most celebrated FZ parafermionic theories are constructed by studying the (degenerate) representations of this algebra. In our notations, these theories are referred as the $\mathbb{Z}_k^{(2)}$ models. The FZ theories are unitary CFTs which describe the critical points of the $\mathbb{Z}_k$ self-dual spin lattice models \cite{FZ}. The case $k=2$ and $k=3$ describe respectively the critical  Ising and three-states Potts model. The $\mathbb{Z}_k$ Read-Rezayi wavefunctions coincide with the $P_n^{(k,2)}$ polynomials. It was shown in \cite{BernevigHaldane1}, that the Read-Rezayi states are written in term of a single Jack:
\begin{eqnarray}
P_n^{(k,2)}(\{z_i\})&=&J^{-k-1}_{\lambda}(\{z_i\})\nonumber \\ \lambda=[\underbrace{N_{\phi},N_{\phi},..}_{\text  k -\mbox{times}}, \underbrace{N_{\phi}-2,N_{\phi}-2,..}_{\text k -\mbox{times}},&\cdots&, \underbrace{2,2,..}_{\text  k -\mbox{times}}];\quad N_{\phi}=\frac{2(n-k)}{k}\label{RRSingleJack}
\end{eqnarray}

\section{Non-unitary and unitary $\mathbb{Z}_{k}^{(r)}$ theories for $r>2$}

Although the study of the FZ theory generalizations has been at the center of an intense research, a complete classification of parafermionic associative solutions for arbitrary $r$ and $k$ is not yet known. 

We present below a series  of $\mathbb{Z}_{k}^{(r)}$ algebras corresponding to particular $k$ and $r$ . The corresponding current correlation functions, and thus the properties of the polynomials $P_{n}^{(k,r)}$ are then discussed.

\subsection{Non-unitary $WA_{k-1}(k+1,k+r)$ models and $\mathbb{Z}_{k}^{(r)}$ theories.}
\label{WAt}
There exists a series of associative solutions which have a fixed value of the central charge and can be related to the so-called $WA_{k-1}$ theories. This is a family of CFTs, the $WA_{k-1}(p,q)$, which is naturally associated to the simple Lie algebra $A_{k-1}$ and indexed by two integers $p$ and $q$\cite{Wtheory}. The  $WA_{k-1}(p,q)$ theories are constructed from the study of the representations of the $WA_{k-1}$ algebras which are generated by a set of chiral currents $\{W_s\}$ ($s=2,\cdots, k$) with spin $s$ ($W_2$ being the stress-energy tensor $T$). The simplest example of $WA_{k-1}$ algebra is the $WA_1$ one which coincides with the Virasoro algebra; the associated theories $WA_{1}(p,q)$ correspond then to the minimal models $M(p,q)$. The models $WA_{k-1}(p,p+1)$, with $p=k+1,k+2,\cdots$ are the unitary representations of the $WA_{k-1}$ algebra. 

As mentioned in the introduction, the first theory in each unitary sequence, corresponding to $p=k+1$ coincides with the FZ parafermions, $\mathbb{Z}_k^{(2)}=WA_{k-1}(k+1,k+2)$. 
   
The model $WA_{k-1}(k+1,k+r)$ is non-unitary for $r>2$ and has a negative central charge $c^{W}(k,r)$\cite{Wtheory}:
\begin{equation}
c^{W}(k,r)=(k-1)\left(1-\frac{k(r-1)^2}{k+r}\right)
\label{c_WA}
\end{equation} 

Let's first consider the $k=2$ case with an arbitrary $r$. The $\mathbb{Z}_2$ symmetry is realized by one single conserved current $\Psi_1(z)=\Psi(z)$ with possible dimension $\Delta=r/4$. 
For $r=2$ ($\Delta=1/2$), $\Psi(z)$ is a free-fermion field and the associated function $P_n^{(2,2)} (\{z_i\})$ describes are Moore-Read states, see section (\ref{FZt}). 
For $r=3,4\cdots$,  it has been observed in \cite{JacobMathieu1,JacobMathieu2,Ravanini} that the non-unitary minimal models $M(3,2+r)$ ($=WA_{1}(3,2+r)$) present an operator in their Kac table (more specifically the $\phi_{1,2}$ operator in the standard notation) whose fusion realizes the $\mathbb{Z}_2^{(r)}$ algebra with central charge $c=c^W(2,r)=r(5-2 r)/(2+r)$.
In particular, one can show that for $c=c^{W}(2,r)$ the $P_{n}^{(2,6)} (\{z_i\}) $ satisfy Eq.(\ref{CS}) with parameter $\alpha=-3/(r-1)$. This can be traced back to the fact that the operator $\Psi(z)=\phi_{1,2}$ satisfies a second level null vector condition \cite{diFrancesco}.
This means that the polynomials $P_{n}^{(2,r)}$ can be written in term of a single Jack:
\begin{eqnarray}
&\quad P_{n}^{(2,r)} (\{z_i\}) = J_{\lambda}^{-3/(r-1)} (\{z_i\})&\nonumber \\
& \lambda=[N_{\phi},N_{\phi},N_{\phi}-r,N_{\phi}-r,\cdots,r,r]&\label{k2_singlejack} \\
&N_{\phi}=r(n-2)/2;\quad c=\frac{r(5-2r)}{2+r}&\nonumber
\end{eqnarray}

 Note that the  above partition $\lambda$ corresponds to the one defined in Eq.(\ref{only_admissible_lambda}) with $k=2$. In \cite{Readnu1} it was shown that   the states with $r > 2$ cannot describe a gapped topological phase. In particular  the state $P_n^{(2,3)} (\{z_i\})$ corresponding to the so-called Gaffnian state \cite{Gaffnian} is not a good candidate for describing gapped phase. 
%%The argument of  \cite{Readnu1} is based on the fact  that the theory of the edge  must be a unitary CFT and the  corresponding central charge is  equal to the effective central charge $c_{eff}$. The  $c_{eff}$ is defined by  counting  the edge excitations and is $c_{eff}=r/(r+2)<1$ for the minimal models $M(3,2+r)$. The value of $c_{eff}$ can also been  computed by using the Jacks theory \cite{BernevigHaldane1,BernevigHaldane2}.  The only unitary theories with $c<1$ are the unitary minimal models $M(m,m+1)$ with $c=1-6/m(m+1)$, $m=3,4,\cdots$. This means that, among the states $P_{n}^{(2,r)} (\{z_i\}) $ defined in Eq.(\ref{k2_singlejack}), the only possible candidates for describing  gapped phase are the ones with $r$ such that $c_{eff}=1-6/m(m+1)$.

Now, consider the case $r=3$ and $k$ taking arbitrary integer values.  It was proved in \cite{JacobMathieu2} that there is an associative algebra $\mathbb{Z}_k^{(3)}$  where the central charge $c$ is fixed to the value $c=c^{W}(k=3,r)$. The corresponding model is shown to be equivalent to the $WA_{k-1}(k+1,k+3)$ one, $\mathbb{Z}_k^{(3)}=WA_{k-1}(k+1,k+3)$.

In general, analogously to the case of the operator $\phi_{1,2}$ appearing in the minimal models $M(3,2+r)$,  a set of  operators $\Phi_{q}^{r}$ of dimension $\Delta_q$ given in Eq.(\ref{dim_ass}) and  realizing a parafermionic fusion $[\Phi_{q}^{r}] \times [\Phi_{p}^{r}]=[\Phi_{q+p}^{r}]$ can be found  the Kac table of the $WA_{k-1}(k+1,k+3)$\cite{Ardonne}theory.  The associativity of the algebra naturally derives from the fact that the conformal blocks  of the $WA_{k-1}(k+1,k+3)$ are well defined.
This means that  there exists an $\mathbb{Z}_k^{(r)}$ algebra with central charge $c=c^{W}(k,r)$ and the corresponding CFT coincides with the $WA_{k-1}(k+1,k+r)$ theory. Moreover, on the basis of the results (\ref{RRSingleJack}) and (\ref{k2_singlejack}) where the polynomials $P_{n}^{(k,r)} (\{z_i\})$ are single Jacks, the general conjecture is that, for $k+1$ and $r-1$ coprimes,  \cite{FJMM, BernevigHaldane2}:
\begin{eqnarray}
&\quad P_{n}^{(k,r)} (\{z_i\}) = J_{\lambda}^{-(k+1)/(r-1)} (\{z_i\})\nonumber \\ 
&\lambda=[\underbrace{N_{\phi},N_{\phi},..}_{\text  k -\mbox{times}}, \underbrace{N_{\phi}-r,N_{\phi}-r,..}_{\text k -\mbox{times}},\cdots ,\underbrace{r,r,..}_{\text  k -\mbox{times}}]&\label{k_singlejack}\\
&N_{\phi}=r(n-k)/k;\quad c=c^{W}(k,r)&\nonumber
\end{eqnarray}
The above  conjecture has been recently verified in \cite{BernevigW} where the central charge, the quasi-hole operator scaling dimensions and the associated fusion rules \cite{Ardonne} were directly obtained from the ground-state (\ref{k_singlejack}) and the quasi-hole wavefunctions: the corresponding results were shown to match with the $WA_{k-1}(k+1,k+r)$ theory.

We stress that for some $r>2$,  there exists $\mathbb{Z}_k^{(r)}(c)$ associative algebras where the central charge $c$ is not fixed but remains a {\it free} parameter. For values of $c=c^{W}(k,r)$, these parafermionic algebras include the abovementioned cases,  $\mathbb{Z}_k^{(r)}(c=c^{W}(k,r))=WA_{k-1}(k+1,k+r)$. Mostly interesting is that the more general $\mathbb{Z}_k^{(r)}(c)$ algebras admit a sequence of unitary CFT for $c$ taking a discrete series of values.

Below we discuss  in details these theories and we compute the corresponding $P_{n}^{(k,r)}$ polynomials. Moreover, by using the techniques developed in \cite{BernevigRegnault}, we give  the expansion of these polynomials on a particular Jacks basis.

\subsection{Unitary $WB_{(r-2)/4}$ models and $\mathbb{Z}_{2}^{(r)}$ theories}

Let's consider in more detail the parafermionic theories based on the $\mathbb{Z}_2$ symmetry which are related to the $k=2$ clustering symmetric polynomials $P_{n}^{(2,r)}$.

 The sequence of values $r=4 l$ ($l=1,2\cdots$) is  special as the dimension of the associated current $\Psi(z)$ is integer, $\Delta=l$. The theory has a continuous symmetry and the parafermionic algebra loses  its main characteristic to realize a discrete symmetry.  Also, the case with odd $r=2l+1$  is special from the $\mathbb{Z}_2$ symmetry point of view:  the field $\Psi(z)$ has dimension $\Delta=(2l+1)/4$ and its correlator functions have branch points instead of a simple monodromy as required by the $\mathbb{Z}_2$ symmetry. 
%%We make the following remark : the correlators $G(z)\hat{=}<\Psi(0)\Psi(1)\Psi(z)\Psi(\infty)>$, for $\Delta=(2l+1)/4$,  satisfy the following boundary conditions:
%%
%%\begin{eqnarray}
%%\lim_{z\to 0} \;\;\sqrt{z}\; G(z)&=&-\lim_{z\to 1}\;\; \sqrt{z-1}\;G(z)\label{boundary}
%%\end{eqnarray}
%%
%%which means that the polynomial $\sqrt{z(z-1)}G(z)$ has odd parity for $z\to 1-z$ (note that this is not inconsistent with the symmetry $G(z)\to G(1-z)$ {\bf since} one has to consider two Riemann sheets).

We focus our attention on the sequence $r=2,6,10,\cdots$.  This sequence of values is the most ``natural'' for the $\mathbb{Z}_2$ symmetry point of view. In fact, the half-integer dimensions of the fields $\Psi(z)$ keeps the fermionic nature of this fields (i.e. the $\Psi$ correlators are antisymmetric).
The $\mathbb{Z}_2$ algebra with $r=6$ ($r=2$ is the FZ theory) coincides with a very well studied  chiral algebra, the so-called $N=1$ superconformal algebra.
In this case, the dimension of the $\mathbb{Z}_2$ conserved current is $\Delta_{1}=3/2$ and it generates the so-called $N=1$ superconformal algebra.

For increasing $r=10,14,\cdots$ the current fusion presents  more and more singular terms ($r/2$ terms) which have to be specified.  Naturally, this is consistent with the fact that the polynomials $P_{2 k}^{(k,r)}$ have higher degree and thus more monomial coefficients have to be determined. 
For increasing $r$ there are more associativity constraints to be satisfied but, at the same time,  more degrees of freedom: it can indeed occur the possibility that, besides the stress-energy  tensor $T(z)$,  other chiral primary fields appear in the current fusions. To each chiral field entering the current OPE  it corresponds a coupling constant which can be tuned to satisfy the additional associativity requirements. Naturally the enclosure of the algebra should then be verified.
For $r=4l+2$ ($\Delta=l+1/2$), $l=0,1,\dots$, it turns out that, in order for  the central charge to be a free parameter,  the parafermionic  algebras have to also contain $l-1$ bosonic fields of dimensions $\Delta=4,6,\cdots,2l$. These algebras are known as $WB_l$ algebras and their representation are naturally associated to the simple Lie algebra $B_l$.

 We stress that the fusion of a field with Abelian monodromy can contain more than one  primary field in its OPE. Consider for instance the case $r=10$ which will be discussed in more detail below. Although the $\mathbb{Z}_2$ current $\Psi$ presents in its fusion  another primary operator $W$ of dimension $4$, $\Psi \Psi\to I+ W$,  the $\Psi$ field has Abelian monodromy: this can be directly seen on the associated functions $P_{4}^{(2,10)}$ which is a  simple polynomial. An operator has a non-Abelian monodromy when, in its fusion, there are primary fields whose dimensions do not differ by integers. The typical example is the spin operator $\sigma$ in the Ising model ($\mathbb{Z}_2^{(2)}$ in our notations) with the fusion $\sigma\sigma \to I+\varepsilon$ where $\Delta_{I}=0$ and $\Delta_{\varepsilon}=1/2$.

Below we present in detail the $r=6$ and $r=10$ cases.

\subsubsection{$r=6$: Superconformal theories}
\label{k2r6}

Here we are interested in computing the symmetric polynomial $P_{4}^{(2,6)}$.
In this case the dimension of the $\mathbb{Z}_2$ conserved current is $\Delta_{1}=3/2$ and generates the so-called $N=1$ superconformal algebra. This algebra, which we denote $\mathbb{Z}_{2}^{6}(c)$, is defined by the fusion:
\begin{eqnarray}
\Psi(z) \Psi(0)=\frac{1}{z^{3}}\left( 1+z^2 \frac{3}{c}T(0)\right)
\label{current_algebra_super}
\end{eqnarray}
The unitary sequence of the $N=1$ superconformal algebra (\ref{current_algebra_super})  has a parametrized central charge :
\begin{equation}
c=\frac{3}{2}\left(1-\frac{8}{m(m+2)} \right) \quad m=3,4\cdots
\label{u_c_super}
\end{equation}
 
For $c=-21/4$ we have seen that $\mathbb{Z}^{(6)}_{2}(-21/4)=M(3,8)$ and the function $P_{4}^{(2,6)}$ is given in Eq.(\ref{k2_singlejack}) in terms of a single Jack:
\begin{equation}
c=-\frac{21}{4}:\quad P_{4}^{(2,6)} (\{z_i\}) = J_{[6,6]}^{-3/5} (\{z_i\})
\label{super_4_singlejack}
\end{equation}

Now we can show explicitly that this is not true for general $c$. 

Let us first consider the behavior of the function $P_{4}^{(2,6)}(z_1,z_2,z_3,z_4)$ when two particles approach at the same point. Using the fusion rule (\ref{current_algebra_super}) 
in the expression (\ref{CFT_FQH}), one has:
\begin{equation}
\lim_{z_2\to z_1}P_{4}^{(2,6)}(z_1,z_2,z_3,z_4)\sim (z_1-z_3)^3(z_2-z_3)^3(z_1-z_4)^3(z_2-z_4)^3\left(1+\frac{3}{c}(z_1-z_2)^2\langle T(z_1)\Psi(z_2)\Psi(z_3)\rangle \right)
\end{equation}
One can observe in the above equation that the function $P_{4}^{(2,6)}(z_1,z_2,z_3,z_4)$ contains a $1/c$ term which vanishes whenever two particles are at the same point. %Consider two homogeneous polynomials in the $n=4$ variables which both satisfy the HW and LW conditions and of the same degree as $P_{4}^{(2,6)}$. We denote this polynomials as $P_{n=4}^{(2,6)(1)} (\{z_i\})$ and $P_{n=4}^{(2,6)(2)} (\{z_i\})$. 
We can write the $P_{4}^{(2,6)}$ polynomial as :
\begin{equation}
P_{4}^{(2,6)} (\{z_i\})=P_{4}^{(2,6)(1)} (\{z_i\})+\frac{1}{c} P_{4}^{(2,6)(2)} (\{z_i\}).
\label{gen_exp_k2r6}
\end{equation}
where $P_{4}^{(2,6)(1)} (\{z_i\})$ and $P_{4}^{(2,6)(2)} (\{z_i\})$ are two homogeneous polynomials in the $n=4$ variables which both satisfy the (HW) and (LW) conditions and are of the same degree as $P_{4}^{(2,6)}$. Note that in the above equation we have excluded the possibility of other terms (for instance $1/c^2$ terms). This is because the algebra (\ref{current_algebra_super}), which does not contain $1/c^2$ terms,  fixes all the singular terms in the fusion $\Psi \Psi$ and thus determines completely the correlation functions. 
 This is also consistent with the fact that, given $N_{\phi}=6$ and $n=4$, the number of independent polynomials dominated by the corresponding root partition and satisfying (HW) and (LW) conditions is $2$. In general the total number of independent homogeneous polynomials dominated by a given root partition can be numerically computed solving the linear equations (HW) and (LW) in the so-called squeezed basis (see e.g. \cite{regnault09}).

 From the expression (\ref{CFT_FQH}) and the fusion rule (\ref{current_algebra_super}) one can easily verify that $P_{4}^{(2,6)(1)} (\{z_i\})$ has to vanish when $3$ particles are at the same point while the term associated to $ P_{4}^{(2,6)(2)} (\{z_i\})$ vanishes when only $2$ particles are at the same point. This means that $P_{4}^{(2,6)(1)} (\{z_i\})$ and $P_{4}^{(2,6)(2)} (\{z_i\})$ belong respectively to the spaces $I^{(2,2,n)}$ and $I^{(1,2,n)}$, see section (\ref{Jacks}). 
On the basis of these considerations and of (\ref{super_4_singlejack}), we obtained the following expansion for the polynomial $P_{4}^{(2,6)}$:
\begin{equation}
P_{4}^{(2,6)} (\{z_i\}) = J_{[6,6]}^{-3/5} (\{z_i\})  +\frac{3(21+4c)}{14 c} J_{[6,4,2]}^{-2} (\{z_i\})  
\label{super_4_jack}
\end{equation}

The form of the expression (\ref{super_4_jack}) is very suggestive: indeed the second term in the r.h.s of the above equation is a kind of "healing term" which keeps the  $k=2$ clustering conditions and makes possible, for the values of $c$ given in (\ref{u_c_super}), to associate the  $P_{4}^{(2,6)}$ polynomials to unitary CFTs.
Note that $J_{[6,4,2]}^{-2}$ is exactly the (bosonic) Laughlin ground state at filling $\nu=1/2$:
\begin{equation}
J_{[6,4,2]}^{-2}(z_1,z_2,z_3,z_4)=\prod_{1\leq i<j\leq 4}(z_i-z_j)^2
\end{equation}

We will see  below how the result in (\ref{super_4_jack}) generalizes to the case of $r=10$.

\subsubsection{$r=10$: $WB_2$ theories}
\label{wb2_four}

Let us consider now the case in which the $\mathbb{Z}_2$ field $\Psi(z)$ has dimensions $\Delta_1=5/2$, 
corresponding to $r=10$ (\ref{dim_ass}). 

The associativity of the chiral algebra (\ref{para_algebra1})-(\ref{para_algebra2}) for arbitrary $c$ requires the introduction of an extra field $W(z)$ with dimension $\Delta_{W}=4$.  The explicit form of such algebra is \cite{Ahn}:
\begin{eqnarray}
\Psi(z)\Psi(w)&=&\frac{1}{(z-w)^5}+\frac{1}{(z-w)^3}\frac{5}{c} T(w)+\frac{5}{2c(z-w)^2}\partial T(w)+\cdots\nonumber\\&+&\frac{1}{z-w}\left(\frac{135}{2c(22+5c)}\Lambda(w)+\frac{15(c-1)}{4c(22+ 5c)}\partial^2 T (w)+\gamma W(w)\right).
\label{wb2algebra}
\end{eqnarray}
In the OPE (\ref{wb2algebra}) $\Lambda(w)$ is a descendant of the identity operator and can be expressed in terms
of the stress energy operator as $\Lambda(w)=:TT:(w)=1/(2\pi i)\oint_{C_w} d z (z-w)^{-1} T(z) T(w)$. 

The algebra constant $\gamma$ is again fixed by associativity requirements to be:
\begin{equation}
\gamma=\frac{75(13+14c)}{c(22+5c)}
\label{wb2_c}
\end{equation} 
The fusion (\ref{wb2algebra}) is a part of the $WB_2$ algebra which is defined by the fusion of the field $W(z)$ as well \cite{Ahn}, and the corresponding unitary sequence have central charge:
\begin{equation}
c=\frac{5}{2}\left(1-\frac{12}{m(m+1)} \right) \quad m=4,5\cdots
\label{central_wb2}
\end{equation} 

We have computed in Appendix (\ref{AppendixWb2})  the four$-$point correlation function $<\Psi(z_1)\Psi(z_2)\Psi(z_3)\Psi(z_4)>$. We have found that the corresponding polynomial $P_{4}^{(2,10)}$ can be written as:
\begin{equation}
P_{4}^{(2,10)} (\{z_i\})=J_{[10,10]}^{-1/3} (\{z_i\})  +\frac{25(22+5c)}{44 c} P^{(2)}(\{z_i\}) 
\label{wb2_4_jack}
\end{equation}
where
\begin{equation}
P^{(2)}(\{z_i\})=J_{[10,8,2]}^{-2} (\{z_i\})-\frac{3}{10} J^{-2}_{[10,7,3]}+\frac{7}{39}J^{-2}_{[10,6,4]}+\frac{25}{546} J^{-2}_{[9,7,3,1]}-\frac{125}{5148}J^{-2}_{[9,6,4,1]}+\frac{5}{567}J^{-2}_{[8,6,4,2]}
\label{p2_wb2}
\end{equation}
As in the case of the superconformal theory, $k=2,r=6$ see Eq.(\ref{super_4_jack}), one can notice that $P_{4}^{(2,10)}$ decomposes into two independent polynomials:
\begin{equation}
P_{4}^{(2,6)} (\{z_i\})=P_{4}^{(2,10)(1)} (\{z_i\})+\frac{1}{c} P_{4}^{(2,10)(2)} (\{z_i\}).
\label{gen_exp_k2r10}
\end{equation}
where $P_{4}^{(2,10)(1)} (\{z_i\})=J_{[10,10]}^{-1/3} (\{z_i\})+125/44 P^{(2)}(\{z_i\}) $ and $P_{4}^{(2,10)(2)} (\{z_i\})=25/44  P^2(\{z_i\})$. These two polynomials are homogeneous states on the sphere,i.e. satisfy the HW and LW conditions,  but with different clustering properties: $P_{4}^{(2,10)(1)} (\{z_i\})$ vanishes  when there is at least $3$ particles occupying the same position ($P_{4}^{(2,10)(1)} (\{z_i\})\in I^{(2,2,4)}$) while  $P_{4}^{(2,10)(2)} (\{z_i\})$ is zero when  as soon as two particles are at the same position ( $P_{4}^{(2,10)(2)} (\{z_i\})\in I^{(1,2,4)}$). From the expression (\ref{p2_wb2}), one can see that $P_{4}^{(2,10)(2)} (\{z_i\})$ can be written in terms of Laughlin states at $\nu=1/2$ with $4$ quasiholes added.

Another interesting remark is that, although the algebra (\ref{wb2algebra}) presents $1/c^2$ terms, the four$-$point correlation function (or equivalently $P_{4}^{(2,10)}$)  does not. This can be traced back again to the associativity requirements: to satisfy these, the constant $\gamma$, defined in Eq.(\ref{wb2_c}),  has to be fixed in such way to cancel the $1/c^2$ terms. 
As observed in section (\ref{k2r6}), this is also consistent with the fact that for $N_{\phi}=10$ and $n=4$, the dimension of independent polynomial satisfying (HW) and (LW) conditions is $2$.

 \subsection{Unitary $\mathbb{Z}_{k}^{(4)}$ theories}
 For $r=4$  it was shown in \cite{FZ} that there exists a parafermionic algebra   which is associative for arbitrary $c$.  The unitary sequence for the  $\mathbb{Z}_3^{(4)}(c)$ model has been investigated in \cite{FZ43}.
  Further developments on the general model $\mathbb{Z}_k^{(4)}(c)$  were presented in \cite{Furlan}, while the explicit construction of their unitary representations has been analyzed in depth in \cite{DotsenkoRaoul}.
  For $k=3$ and $r=8$ an associative solution  has been constructed in \cite{DotsenkoRaoul2}. The representation theory of the $\mathbb{Z}_{3}^{(8)}(c)$ algebra is not known.
  
\subsubsection{$k=3$, $r=4$}
In the previous sections we have analyzed  the wavefunctions associated to some $\mathbb{Z}_2$ ($k=2$) parafermionic current algebras. 
We now move to the symmetry $\mathbb{Z}_3$ and to the wavefunctions satisfying $k=3$ clustering properties. In the case $\mathbb{Z}_3$ we have two parafermions $\Psi(z)$ and $\Psi^{\dagger}(z)$ with the same dimension $\Delta=r/3$. Analogously  to the case $\mathbb{Z}_2$ symmetry,   the values of $r=3,6,9,\cdots$ are not natural for the $\mathbb{Z}_3$ symmetry: the currents have an integer dimension and the symmetry of the theory is a continuous one.

The next interesting case after $r=2$ is $r=4$, corresponding to $\Delta=4/3$. The corresponding  algebra reads explicitly:
\begin{eqnarray}
\Psi(z) \Psi(w)&=&\frac{\gamma}{(z-w)^{4/3}}\left(\Psi^{\dagger}(w)+(z-w)\partial_w \Psi^{\dagger}(w)+\cdots\right)
\label{para_algebra1_z3}\\
\Psi(z) \Psi^{\dagger}(w)&=&\frac{1}{(z-w)^{8/3}}\left(1+(z-w)^2\frac{8}{3 c} T(w)+\cdots \right)
\label{para_algebra_z3}
\end{eqnarray}
where $\gamma=4/3\sqrt{2/c-1/4}$ is a constant fixed by associativity requirements and $c$ is free.
The unitary sequence of the $\mathbb{Z}_3^{(4)}(c)$ model has central charge \cite{zomo32} : 
\begin{equation}
c=2\left( 1-\frac{12}{m(m+4)}\right)\quad m=3,4\cdots
\end{equation}
The first theory of the unitary series ($m=3$)  coincides with the unitary minimal model $M(6,7)$ with central charge $c=6/7$ which describes the tri-critical behavior of the three-states Potts model.

We are now interested to the function:
\begin{equation}
P_{6}^{(3,4)} (\{z_i\}) =  \langle \Psi(z_1) \Psi(z_2) \Psi(z_3)\Psi(z_{4}) \Psi(z_{5})  \Psi(z_{6})\rangle \prod_{i<j} \left(z_i-z_j \right)^{4/3}. 
\label{p6_3_4}
\end{equation}

We have computed explicitly (for details see Appendix \ref{App3}) the function (\ref{p6_3_4}). We found the following expansion in term of Jacks :
\begin{equation}
P_{6}^{(3,4)} (\{z_i\})= J_{[4,4,4]}^{-4/3}(\{z_i\}) + \frac{4(40+7c)}{45 c}J_{[4,4,2,2]}^{-3}(\{z_i\}) \\
\label{p6_3_4_ja} 
\end{equation}
For $c=-40/7$ one recovers the result of \cite{BernevigHaldane1} and conjectured to correspond to the correlators of the non-unitary theory $WA_{2}(3,7)$. For $c=8/5=2\times 4/5$, the  $\mathbb{Z}_{3}^{(2)}$ theory factorizes into two copies of  $\mathbb{Z}_3^{(r=2)}$ CFT.  We have thus verified that, for this value of $c$ :
\begin{equation}
P_{6}^{(3,4)} (\{z_i\})=\left[P_{6}^{(3,2)} (\{z_i\})\right]^2=\left[J^{-4}_{[2,2,2]} (\{z_i\})\right]^2 \quad \mbox{for}\;c=8/5.
\end{equation}
Analogously  to the case $k=2$, one can observe that  the function $P_{6}^{(3,4)}$ admits a decomposition into two terms: 
\begin{equation}
P_{6}^{(3,4)} (\{z_i\})=P_{6}^{(3,4)(1)} (\{z_i\})+\frac{1}{c} P_{6}^{(3,4)(2)} (\{z_i\}).
\label{gen_exp_k3r4}
\end{equation}
The term $P_{6}^{(3,4)(1)} (\{z_i\})=J_{[4,4,4]}^{-4/3}+28/45 J_{[4,4,2,2]}^{-3}(\{z_i\}) $ satisfies  the $k=3-$clustering properties and the $1/c$ term $ P_{6}^{(3,4)(2)} (\{z_i\})=(32/7) J_{[4,4,2,2]}^{-3}$  satisfies $k=2$ clustering properties.

 The "correction" term $J_{[4,4,2,2]}^{-3}$ describes the Moore-Read states and allows to move from the non-unitary theory associated to the Jack $J_{[4,4,4]}^{-4/3}$ to a unitary one.

\subsubsection{$k=5$, $r=4$}

Another interesting example is that of $k=5,r=4$. In that case $r-1$ and $k+1$ are not coprimes, and the corresponding W-theory $WA_4(3,9)$ is not well defined. However the parafermionic theory $\mathbb{Z}_{5}^{(4)}$ is a perfectly valid CFT, and the central charge $c$ is a free parameter. The unitary sequence of the $\mathbb{Z}_3^{(4)}$ model has central charge \cite{DotsenkoRaoul} :
\begin{equation}
c=4\left( 1-\frac{15}{m(m+2)}\right)\quad m=4,5\cdots
\end{equation}

The second theory of the unitary series ($m=5$) with  central charge $c=\frac{16}{7}$, coincides with two copies of the FZ $k=5$ parafermions. We are now interested in the function :
\begin{equation}
P_{10}^{(5,4)} (\{z_i\}) =  \langle \Psi(z_1) \Psi(z_2) \dots \Psi(z_{10}) \rangle \prod_{i<j} \left(z_i-z_j \right)^{4/5}. 
\label{p10_5_4}
\end{equation}

We have computed explicitly (for details see Appendix \ref{App5}) the function (\ref{p10_5_4}). The following expansion in term of Jacks was obtained:
\begin{equation}
P_{10}^{(5,4)} = \left( J_{[2,2,2,2,2]}^{-6} \right)^2 + \frac{128}{25} \left( \frac{1}{c} - \frac{7}{16}\right) \left(  J_{[4,4,4,4,2,2]}^{-5} -\frac{3}{8} J_{[4,4,4,3,2,2,1]}^{-5} + \frac{18}{7} J_{[4,4,4,2,2,2,2]}^{-5} + \frac{3}{11} J_{[4,4,3,3,2,2,1,1]}^{-5}  \right)
\label{p4_k5r4}
\end{equation}
Note that , as we already mentioned before, we cannot compare the Eq. (\ref{p4_k5r4}) with the Eq.(\ref{k_singlejack}). This can be seen directly from the fact that the Jack $J^{\alpha}_{[4,4,4,4]}$ has a pole for $\alpha=-(k+1)/r-1=-2$.

\section{General structure for $n$-point functions $P_{n=2k}^{(k,r)}(\{z_i\})$}
In the previous sections we have computed explicitly the $P_{n=2k}^{(k,r)}(\{z_i\})$ symmetric functions generated by various generalizations of parafermionic theories and  we have discussed in details their clustering properties. These properties are made manifest by the expansions in terms of Jacks we presented.  In particular,  we have seen in Eqs  (\ref{gen_exp_k2r6}),(\ref{gen_exp_k2r10}),(\ref{gen_exp_k3r4}) and in the Eq.(\ref{p4_k5r4})  that the $P_{2k}^{(k,r)}(\{z_i\})$ is written in terms of two polynomials $P_{2 k}^{(k,r)(1)}(\{z_i\})$, $P_{2k}^{(k,r)(2)}(\{z_i\})$:
\begin{equation}
P_{2k}^{(k,r)}(\{z_i\})=P_{2k}^{(k,r)(1)}(\{z_i\})+\frac{1}{c}P_{2k}^{(k,r)(2)}(\{z_i\})
\label{summ_res}
\end{equation}
characterized by the following properties:
\begin{itemize}
\item $P_{2k}^{(k,r)(1)}(\{z_i\})$ and $P_{2k}^{(k,r)(2)}(\{z_i\})$  are symmetric functions of $2k$ variables which  satisfy LW and HW conditions
\item  $P_{2k}^{(k,r)(1)}(\{z_i\}) \in I^{(k,2,n=2k)}$, $P_{2 k}^{(k,r)(2)}(\{z_i\}) \in I^{(k-1,2, n= 2k)}$
\end{itemize}
For each $\mathbb{Z}_{k}$ theory, the $2 k$ particles functions are the building blocks to compute the most general $n$ particles functions $P_{n}^{(k,r)}(\{z_i\})$. 
In fact, using the current OPEs, it is possible to compute recursively the $n-$point correlation functions by relating them to the $n-1$ and $n-2$ ones. In the Appendix \ref{AppendixWb2} we show this procedure for the case $\mathbb{Z}_{2}^{(10)}(c)$.

As we show below, the form of the parafermionic algebra $\mathbb{Z}_{k}^{(r)}(c)$ implies a general structure for the function   $P_{n}^{(k,r)}(\{z_i\})$: it can be expressed in terms of $l$ independent polynomials $P_{l k}^{(k,r)(i)}(\{z_i\})$ ($i=1,\cdots, l$),
\begin{equation}
P_{lk}^{(k,r)}(\{z_i\})=P_{lk}^{(k,r)(1)}(\{z_i\})+\frac{1}{c}P_{lk}^{(k,r)(2)}(\{z_i\})+\frac{1}{c^2}P_{lk}^{(k,r)(3)}(\{z_i\})+\cdots \frac{1}{c^{l-1}}P_{lk}^{(k,r)(l)}(\{z_i\}),
\label{gen_expans}
\end{equation}
which are characterized by the following properties:
\begin{itemize}
\item $P_{lk}^{(k,r)(i)}(\{z_i\})$ , $i=1,\cdots, l$ are symmetric functions all  satisfying the LW and HW conditions (\ref{HW})-(\ref{LW}).

\item $P_{lk}^{(k,r)(1)}(\{z_i\})\in I^{(k,2,n=lk)}$ (see section \ref{Jacks})

\item $P_{lk}^{(k,r)(i)}(\{z_i\})$ $i=2,\cdots,l$ vanishes when $l-i+1$ cluster of $k$ particles are formed:
\begin{eqnarray}
P_{lk}^{(k,r)(i)}(\underbrace{z_1=\cdots=z_{k}}_{\text Z_1}, &..&,\underbrace{z_{(l-i-1)k+1}=\cdots=z_{(l-i)k}}_{\text Z_{l-i}},z_{(l-i)k+1}, z_{(l-i)k+2},\cdots)\nonumber \\
&=&\prod_{1=i<j}^{l-i}(Z_{i}-Z_{j})^{k r}\prod_{i=1}^{l-i}\prod_{j\geq(l-i)k+1}(Z_i-z_{j})^r P_{ik}^{(k,r)(i)}\label{ann_clust}\\
P_{lk}^{(k,r)(i)}(\underbrace{z_1=\cdots=z_{k}}_{\text Z_1}, &..&,\underbrace{z_{(l-i-1)k+1}=\cdots=z_{(l-i)k}}_{\text Z_{l-i}},\underbrace{z_{(l-i)k+1}=\cdots=z_{(l-i+1)k}}_{\text Z_{l-i+1}},..)=0\label{gen_clust}
\end{eqnarray}
\end{itemize}

We show now how the form of the $\mathbb{Z}_{k}^{(r)}(c)$ algebra leads to the properties (\ref{gen_expans})-(\ref{gen_clust}). For the sake of clarity, we discuss the case where $k=2$, $r=6$. The polynomial  $P_{n}^{(2,6)}$ (\ref{CFT_FQH}) reads:
\begin{equation}
P_{n}^{(2,6)}(\{z_i\})=\langle \underbrace{\Psi(z_1) \Psi(z_2) \dots   \Psi(z_n)}_{n \textrm{ fields}} \rangle \prod_{i<j}(z_i-z_j)^3. \nonumber 
\end{equation}

Using the OPEs for the parafermionic field $\Psi$ in $\mathbb{Z}_{2}^{6}(c)$, one can derive the following relation for the $n$-point correlation function $\langle \Psi(z_1) \Psi(z_2) \dots   \Psi(z_n) \rangle$ \cite{zamo_infinite,Simon_super}:

%\begin{eqnarray}
%\langle \Psi(z_1)\Psi(z_2)\cdots  \Psi(z_n)\rangle&=&\sum_{\substack{j=2,3\cdots n\\ l,m,p\cdots \neq 1,j;\; l<m<\cdots<p}}\frac{(-1)^{j}}{(z_1-z_j)^5}\langle  \Psi(z_l) \Psi(z_m)\cdots \Psi(z_p)..\rangle+\nonumber \\
%&&+\frac{3}{c}\sum_{\substack{j=2,3\cdots n\\ l,m,p\cdots \neq 1,j;\; l<m<\cdots<p}}\frac{(-1)^{j}}{(z_1-z_j)^3}\langle  T(z_j) \Psi(z_l) \Psi(z_m)\cdots \Psi(z_p)..\rangle+\nonumber 
%\end{eqnarray}

%Now, using the relation :
%\begin{equation}
%\langle T(z)\Psi(z_1) \Psi(z_2) \cdots \Psi(z_n) \rangle = \sum_{i}\left( \frac{3/2}{(z-z_i)^2} + \frac{\partial_i}{z-z_i}\right) \langle \Psi(z_1) \Psi(z_2) \cdots \Psi(z_n) \rangle 
%\end{equation}

%\begin{eqnarray}
%P_{n}^{(k=2,r=6)} = \sum_{m=2}^{n} \prod_{i \neq 1,m}(z_{1i})^3(z_{mi})^3 \left(1+\frac{3}{c}(z_{1m})^{2} \left( \sum_{i\neq1,m}\left( \frac{3/2}{(z_{1i})^2} + \frac{\partial_i}{z_{1i}}\right) - \sum_{\substack{i<j\\ i,j \neq 1,m}} \frac{3}{(z_{1i})(z_{1j})} \right) \right) P_{n-2}^{k=2,r=6}\left(z_2, \cdots z_{m-1},z_{m+1},\cdots z_n \right) 
%\label{es2_k2r6}
%\end{eqnarray}

\begin{eqnarray}
\langle \prod_{i=1}^n\Psi(z_i)\rangle = \sum_{m=2}^{n} \frac{(-1)^m}{(z_{1}-z_{m})^3} \left(1+\frac{3}{c}(z_{1}-z_{m})^{2}  \sum_{i\neq1,m}\left( \frac{3/2}{(z_{1}-z_{i})^2} + \frac{\partial_i}{z_{1}-z_{i}}\right)  \right)  \langle \prod_{\substack{i=1\\ i \neq 1,m}}^n \Psi(z_i) \rangle 
\label{es2_k2r6}
\end{eqnarray}

From this recursion relation it is clear that the $n$-point function $P_{n}^{(2,6)}$ admits an expansion in powers of $1/c$ :

\begin{equation}
 P_{n=2l}^{(2,6)} = \sum_{i=1}^{l} \frac{1}{c^{i-1}}  P_{n=2l}^{(2,6)(i)}
\end{equation}

where the polynomials $P_{n=2l}^{(k=2,r=6)(i)}$, $i=1 \cdots l$ are independent of $c$. Moreover they have to satisfy the LW and HW conditions, and they must obey the same clustering properties as $P_{n}^{(2,6)}$. They all vanish when one cluster of $k+1=3$ particles is formed, and :

\begin{equation}
P_{n=2l}^{(2,6)(i)} (z_1=Z_1,z_2=Z_1,z_3\cdots z_n) = \prod_{i=3}^{n} (Z_1-z_i)^6 P_{n=2(l-1)}^{(2,6)(i)} (z_3,z_4,\cdots z_{n})
\end{equation}

Iteratively, forming $p$ clusters of 2 particles $z_{2m-1}=z_{2m}=Z_m$ for $i=1\cdots p$, one gets : 

\begin{eqnarray}
P_{n=2l}^{(2,6)(i)} (\underbrace{z_1,z_2}_{z_1=z_2=Z_1},\underbrace{z_3,z_4}_{z_3=z_4=Z_2},\cdots \underbrace{z_{2p-1},z_{2p}}_{z_{2p-1}=z_{2p}=Z_{p}},z_{2p+1}, \cdots z_n )  & = & \\ \prod_{1\leq i<j \leq p} (Z_i-Z_j)^{12} \prod_{i=1}^p \prod_{j=2p+1}^n (Z_i-z_j)^6 P_{n=2(l-p)}^{(2,6)(i)} (z_{2p+1},\cdots z_{n}) & & \end{eqnarray}

Since $P_{n=2l}^{(2,6)(i)}=0$ for $i\geq l+1$ (except when $l=0$ and $i=0$), it is clear that for $i\geq 2$, $P_{n=2l}^{(2,6)(i)} $ vanishes when $p=l-i+1$ clusters of $k=2$ particles are formed. This proves the results announced in this section for the case $k=2$, $r=6$.

We have verified the form (\ref{gen_expans}) by computing explicitly the $n$-point correlation functions of the $\mathbb{Z}_{2}^{(6)}(c)$ and $\mathbb{Z}_{2}^{(10)}(c)$ theories.
Recently, a compact expression to generate the  $n$-point correlation function of the $\mathbb{Z}_{2}^{(6)}(c)$ theory  has been explicitly  given in \cite{Simon_super}.  It was shown that, analogously to the Ising case ($\Delta_{1}=1/2$), one can express the $P_{2 n}^{(2,6)} (\{z_i\}) $ polynomial in terms  of  the four-point function:
\begin{equation}
\chi(z_1,z_2;z_3,z_4)=(\frac{1}{3}-\frac{1}{c})(z_1-z_3)^3(z_1-z_4)^3(z_2-z_4)^3(z_2-z_3)^3+ \frac{1}{c}(z_1-z_3)^4(z_1-z_4)^2(z_2-z_4)^4(z_2-z_3)^2
\label{chi_simon}
\end{equation}
by:
\begin{equation}
P_{n=2l}^{(2,6)} (\{z_i\}) =\sum_{P}\prod_{1\leq i<j\leq l}\chi(z_{P(2 i-1)},z_{P(2 i)};z_{P(2j -1)},z_{P(2 j)})
\label{gen_simon}
\end{equation}
One can notice that, for the case $n=6$ for instance, the $P_{n}^{(2,6)} (\{z_i\})\sim \sum_{P} \chi \chi \chi$ contains a product of three $\chi$ functions. The term $1/c^3$ coming by expanding this product is antisymmetric and vanishes after  symmetrization. One then recovers the expansion (\ref{gen_expans}).

A highly non-trivial check of (\ref{gen_expans}) is provided by the theory $WB_2$ corresponding to $k=2$ and $r=10$ (see section \ref{wb2_four}). In this case, the corresponding  $\Psi \Psi$ fusion presents singular terms with $1/c^2$ coefficients entering in the computation of correlation functions. We have derived in  Appendix \ref{AppendixWb2} the recursive relation to compute the $n$-point correlation functions. The explicit results obtained for $n=4,6,8$  match with the expansion (\ref{gen_expans}).

Finally we remark that very similar clustering properties as the ones characterizing the polynomials  $P_{n}^{(k,r)(i)}$ defined above  were considered in \cite{BernevigHaldane_gen_clust}: the behavior of symmetric functions $P(z_1,\cdots,z_n)$ which i) vanish when a number $s$ of clusters of $k$ particles is formed and ii) does not vanish when a cluster of $s k-1$ particles is formed were studied   \cite{BernevigHaldane_gen_clust}. Note that this last property is not satisfied by our  $P_{n}^{(k,r)(i)}$ polynomials which all vanish when a cluster of $k+1$ is formed.

\section{Conclusions}
In this paper we considered  polynomials   $P_{n}^{(k,r)}(\{z_i\})$ which are symmetric in the $n$ variables $z_i$  and which correspond to the densest  polynomials  satisfying  $(k,r)$ clustering properties. The interest in these polynomials  is motivated because of their possible connection to  non-Abelian bosonic FQH  systems at filling fraction $\nu=k/r$.  Recent studies have considered  polynomials  $P_{n}^{(k,r)}$ which are expressed in terms of single Jacks and which are given by correlation functions of non-unitary CFTs.  However, there are solid arguments that the states derived by non-unitary CFTs cannot describe gapped phases,

We presented a systematic study of the CFT approach to generate the $P_{n}^{(k,r)}$ polynomials.  As we pointed out,   this problem is directly connected to the classification of parafermionic algebras.  Particular attention  has been given to the possibility of generating the $P_{n}^{(k,r)}$ polynomials  by using unitary CFTs.  We have then considered in details  certain families of  parafermionic algebras  $\mathbb{Z}_{k}^{(r)}(c)$  which have a  central charge $c$ as a free parameter. The $\mathbb{Z}_{k}^{(r)}(c)$ algebras studied here  are known to  present unitary sequence of CFTs. An example is the  case $\mathbb{Z}_{k=2}^{2+4l}$($l=1,2\cdots$), which  coincides  with the $W$ theories associated to the $B_l$ Lie algebras.
 We computed the $\mathbb{Z}_{k}^{(r)}(c)$ correlation functions for different values of $k$ and $r$ and  we gave the explicit expansion of  the associated $P_{n}^{(k,r)}$ polynomials in terms of Jack polynomials.  For particular (negative) values of the central charge $c$, our results coincides with  the single Jack states  studied in previous works and given  by  non-unitary $WA_{k-1}(k+1,k+r)$ theories. For other values of the central charge $c$,  including the sequence of unitary CFTs, we show that the $P_{n}^{(k,r)}$ polynomials can be expanded in a  series of  Jacks  with specific generalized clustering conditions.

At our knowledge,  many of the unitary CFTs discussed in this paper have never been considered before in the study of  non-Abelian FQH states. In a forthcoming work, we will address the essential question of the quasiparticle properties induced by these CFTs and the eventual connection with the approach based on patterns of zeros \cite{Wen2,Wen3}. It is important to remark that the states $P_{n}^{(k,r)}$  are the zero-energy eigenstates of the pseudo-Hamiltonian $\mathcal{P}_{k}^{r}$ but, for $(k=2,r>4)$ and for $(k>3,r>2)$, they are not unique. We did not find an Hamiltonian which admits the $P_{n}^{(k,r)}$ states as  unique ground states. However, S. Simon has informed us that there appears to be a Hamiltonian that produces the $\mathbb{Z}_{3}^{(4)}(c)$
state\cite{Simonrezayi_private}. We mention also that the  $\mathbb{Z}_{3}^{4}(c)$ theory was considered in \cite{Simonrezayi_unp} where the states $\mathcal{P}_{3}^{(4)}$ were compared to exact diagonalizations results of bosons at $\nu=3/4$. The higher overlap were obtained for non-unitary members of the series. 
 It would then be interesting to check the predictions of these CFTs  and to  verify if the unitary functions $P_{n}^{(k,r)}$ we computed can or not be good candidate for describing non-Abelian states.

{\it Acknowlegdgement}: R.S.  thanks N. Cooper for very helpful discussions. He is  particularly grateful to Th. Jolicoeur for very useful advises and support all along this work. N.R. acknowledges B.A Bernevig for numerous discussions and insightful remarks, and N. Read for careful reading. We also aknowledge S.Simon for pointing out his talk at IPAM 2007 and for communicating us some of his unpublished results.

\section{Appendix}

\subsection{Computation of the $\Psi$ correlation functions for $\mathbb{Z}_{2}^{(10)}$}
\label{AppendixWb2}

We show here how to compute the $\Psi$ correlation functions of the $WB_2$ theory by using the OPEs
 (\ref{wb2algebra}) and the fusions between the fields $W$ and $\Psi$:
 \begin{eqnarray}
W(z) \Psi(w)&=&\frac{\gamma}{(z-w)^4} \Psi(w)+\frac{\gamma_{1}}{(z-w)^3}\partial_{w}\Psi(w) +\frac{1}{(z-w)^2}(\gamma_{11}\partial^{2}_{w}\Psi(w)+\gamma_{2}L_{-2}\Psi(w))+\nonumber \\
&&+\frac{1}{(z-w)}(\gamma_{111}\partial^3_{w}\Psi(w)+\gamma_{12}\partial_{w}L_{-2}\Psi(w)+\gamma_{3}L_{-3}\Psi(w)) +\mbox{Regular terms}\label{WP_fusion}
\end{eqnarray}
In the above equation $\gamma$ is the algebra constant given in Eq.(\ref{wb2_c}), the $L_{n}$ are the stress-energy tensor $T(z)$ modes:
\begin{equation}
L_{n} \Psi(w)=\frac{1}{2\pi i}\oint_{C_w} d z (z-w)^{n+1} T(z) \Psi(w),
\end{equation}
and the constants $\gamma_{..}$ are fixed by comparing the conformal transformations of the l.h.s and r.h.s. of Eq.(\ref{WP_fusion}) :
\begin{eqnarray}
\gamma_{1}&=&\frac{4}{5}\gamma \quad \gamma_{11}=\frac{2(8 + c)}{3(25 + 2 c)}\gamma \quad \gamma_{111}=\frac{4(-110 + 10 c^2 + 19 c)}{15(325 + 376 c + 28 c^2)}\gamma \nonumber \\
\gamma_{2}&=&\frac{12}{25+2c}\gamma \quad \gamma_{12}=\frac{8(82 c + 215)}{5(325 + 376 c + 28 c^2)}
\gamma \nonumber \\
\gamma_{3}&=&\frac{-108}{5(14 c+13)} \gamma
\end{eqnarray}

Using the OPEs (\ref{wb2_c}) one has:
\begin{eqnarray}
\langle \Psi(z_1)\Psi(z_2)\cdots  \Psi(z_n)\rangle&=&\sum_{\substack{j=2,3\cdots n\\ l,m,p\cdots \neq 1,j;\; l<m<\cdots<p}}\frac{(-1)^{j}}{(z_1-z_j)^5}\langle  \Psi(z_l) \Psi(z_m)\cdots \Psi(z_p)..\rangle+\nonumber \\
&&+\frac{5}{c}\sum_{\substack{j=2,3\cdots n\\ l,m,p\cdots \neq 1,j;\; l<m<\cdots<p}}\frac{(-1)^{j}}{(z_1-z_j)^3}\langle  T(z_j) \Psi(z_l) \Psi(z_m)\cdots \Psi(z_p)..\rangle+\nonumber \\
&&+\frac{5}{2c}\sum_{\substack{j=2,3\cdots n\\ l,m,p\cdots \neq 1,j;\; l<m<\cdots<p}}\frac{(-1)^{j}}{(z_1-z_j)^2}\partial_{z_j} \langle  T(z_j) \Psi(z_l) \Psi(z_m)\cdots \Psi(z_p)..\rangle+\nonumber\\
&&+\frac{15(-1 + c)}{4 c (22 + 5 c)}\sum_{\substack{j=2,3\cdots n\\ l,m,p\cdots \neq 1,j;\; l<m<\cdots<p}}\frac{(-1)^{j}}{(z_1-z_j)}\partial^2_{z_j} \langle  T(z_j) \Psi(z_l) \Psi(z_m)\cdots \Psi(z_p)..\rangle+\nonumber\\
&&+\frac{135}{2 c (22 + 5 c)}\sum_{\substack{j=2,3\cdots n\\ l,m,p\cdots \neq 1,j;\; l<m<\cdots<p}} \frac{(-1)^{j}}{(z_1-z_j)}\langle  L_{-2}T(z_j) \Psi(z_l) \Psi(z_m)\cdots \Psi(z_p)..\rangle+\nonumber \\
&&+\gamma  \sum_{\substack{j=2,3\cdots n\\ l,m,p\cdots \neq 1,j;\; l<m<\cdots<p}}\frac{(-1)^{j}}{(z_1-z_j)}\langle  W(z_j) \Psi(z_l) \Psi(z_m)\cdots \Psi(z_p)..\rangle \label{psipsi}
\end{eqnarray}
Analogously, from the expansion (\ref{WP_fusion}) we have:
\begin{eqnarray}
\langle  W(z_1) \Psi(z_2) \Psi(z_3)\cdots \Psi(z_n)\rangle&=&\gamma \sum_{\substack{j=2,3\cdots n\\ l,m,p\cdots \neq 1,j;\; l<m<\cdots<p}}\frac{(-1)^{j}}{(z_1-z_j)^4}\langle  \Psi(z_j)\Psi(z_l) \Psi(z_m)\cdots \Psi(z_p)..\rangle+\nonumber \\
&&+\gamma_1 \sum_{\substack{j=2,3\cdots n\\ l,m,p\cdots \neq 1,j;\; l<m<\cdots<p}}\frac{(-1)^{j}}{(z_1-z_j)^3}\partial_{z_j} \langle  \Psi(z_j)\Psi(z_l) \Psi(z_m)\cdots \Psi(z_p)..\rangle+\nonumber \\
&&+\gamma_{11}\sum_{\substack{j=2,3\cdots n\\ l,m,p\cdots \neq 1,j;\; l<m<\cdots<p}}\frac{(-1)^{j}}{(z_1-z_j)^2}\partial_{z_j}^{2}\langle  \Psi(z_j)\Psi(z_l) \Psi(z_m)\cdots \Psi(z_p)..\rangle+\nonumber \\
&&+\gamma_{2}\sum_{\substack{j=2,3\cdots n\\ l,m,p\cdots \neq 1,j;\; l<m<\cdots<p}}\frac{(-1)^{j}}{(z_1-z_j)^2}\langle L_{-2} \Psi(z_j)\Psi(z_l) \Psi(z_m)\cdots \Psi(z_p)..\rangle+\nonumber \\
&& +\gamma_{111}\sum_{\substack{j=2,3\cdots n\\ l,m,p\cdots \neq 1,j;\; l<m<\cdots<p}}\frac{(-1)^{j}}{(z_1-z_j)}\partial_{z_j}^{3}\langle L_{-2} \Psi(z_j)\Psi(z_l) \Psi(z_m)\cdots \Psi(z_p)..\rangle+\nonumber \\
&&+\gamma_{12}\sum_{\substack{j=2,3\cdots n\\ l,m,p\cdots \neq 1,j;\; l<m<\cdots<p}}\frac{(-1)^{j}}{(z_1-z_j)}\partial_{z_j}\langle L_{-2} \Psi(z_j)\Psi(z_l) \Psi(z_m)\cdots \Psi(z_p)..\rangle+\nonumber\\
&&+\gamma_{3}\sum_{\substack{j=2,3\cdots n\\ l,m,p\cdots \neq 1,j;\; l<m<\cdots<p}}\frac{(-1)^{j}}{(z_1-z_j)}\langle L_{-3} \Psi(z_j)\Psi(z_l) \Psi(z_m)\cdots \Psi(z_p)..\rangle \label{Wpsi}
\end{eqnarray}
In order to explicitly compute  the above expression we need the following general result:
\begin{eqnarray}
\langle L_{n}\Phi(z_1) \Phi(z_2)\cdots \Phi(z_n))\rangle&=& \frac{1}{2\pi i}\oint_{C_{z_1}}(z-z_1)^{n+1} \langle T(z) \Phi(z_1) \Phi(z_2)\cdots \Phi(z_n))\rangle=\nonumber \\&&=-\sum_{j=2}^{n}  \frac{1}{2\pi i} \oint_{C_{z_j}}(z-z_1)^{n+1}  \langle \Phi(z_1) \Phi(z_2)\cdots T(z) \Phi(z_j)\cdots \Phi(z_n))\rangle=\nonumber \\
 &&=-\left(\sum_{j=2}^{n}  \frac{(n+1)\Delta_{\Phi}}{(z_j-z_1)^{-n}}+\frac{1}{(z_j-z_1)^{-n-1}}\partial_{z_j}\right)\langle \Phi(z_1)\Phi(z_2)\cdots\Phi(z_n)\rangle
\label{modes_Ln}
\end{eqnarray}
which come from  standard deformation of complex contour integrals and by the primary operator property:
\begin{equation}
T(z)\Phi(w)=\frac{\Delta_{\Phi}}{(z-w)^2}\Phi(w)+\frac{1}{z-w}\partial_{w}\Phi(w)
\end{equation}  
Using Eq.(\ref{psipsi})-(\ref{modes_Ln}) one can evaluate iteratively all the $n-$point correlation functions. With the help of  Mathematica we could exactly evaluate the four and six point correlation functions.

\subsection{Computation of the $2k-$point function for $\mathbb{Z}_{k}^{(r)}$, with $r=4$}

We are interested in the following polynomial :
\begin{eqnarray}
P_{2k}^{(k,r)} (\{z_i\}) & = & \langle \Psi(z_1) \dots \Psi(z_{2k})\rangle \prod_{i<j} \left( z_{ij} \right)^{2(k-1)/k} 
\end{eqnarray}

It is obviously symmetric in $z_i$, and it enjoys the following properties, inherited from the underlying CFT $\mathbb{Z}_{k}^{(r)}$ (cf appendix \ref{AppConf}):

\begin{itemize}
 \item The degree in each variable $z_i$ is $N_{\Phi}= r $ 
 \item It is homogeneous with total degree $\frac{1}{2}n N_{\Phi} =  k r$
 \item It obeys the (LW) and (HW) conditions
 \item \emph{clustering} property : $ P_n^{(k,r)} (z_1=Z,\dots  , z_{2}=Z ,z_{k} = Z, z_{k+1},\cdots,z_n)\propto \prod_{i=k+1}^{n} (Z-z_i)^{r}$

\end{itemize}

One can expand this polynomial in the monomial basis  $\left\{m_{\lambda} \right\}$, where $\lambda$ must obey : $|\lambda| = \sum_i \lambda_i = k r, \quad l(\lambda)\leq n, \quad \lambda_1 \leq r $. Thus $\lambda$ is nothing but a partition of $k r$ that fits in a $2k \times r$ rectangle.

%The generating function for such partitions is the q-binomial coefficient :
%\begin{equation}
%\left[ \begin{array}{c} m+n \\ n \end{array}\right]_q = \frac{(q)_{m+n}}{(q)_{n}(q)_{m}}
%\end{equation} 

%with \begin{equation}(q)_n= \prod_{i=1}^{n}(1-q^i) \quad  \quad \left(  n \geq 0 \right) \end{equation}

%The q-binomial coefficient $\left[ \begin{array}{c} m+n \\ n \end{array}\right]_q$ can be interpreted as a polynomial in $q$ whose coefficient $q^k$ counts the number of distinct partitions of $k$ elements which fit inside an $m \times n$ rectangle. Therefore the dimension of the vectorspace of symmetric homogeneous polynomials of $N$ variables, degree $N_{\Phi}$ in each variable and total degree $D$ is simply the coefficient $q^D$ of the polynomial $ \left[ \begin{array}{c} N+N_{\Phi} \\ N_{\Phi} \end{array}\right]_q $.

\begin{eqnarray}
P_{n=2k}^{(k,r=4)} (\{z_i\}) & = & \sum_{\lambda \leq [\underbrace{r,r,\dots,r}_{k}] } a_{\lambda} m_{\lambda}
\end{eqnarray}

The first terms in this expansion can be obtained by considering the following specialization : 

\begin{eqnarray}
P_{2k}^{(k,r)} (z_1=\dots=z_{k-1}=\infty,z_k=1,z_{k+1}=z,z_{k+2}=\dots=z_{2k}=0)   & \propto &  R_k(z) \label{specialization}
\end{eqnarray}

where
\begin{equation}
 R_k(z) = \langle \Psi^{\dagger}(\infty) \Psi(1) \Psi(z) \Psi^{\dagger}(0)\rangle \left( 1-z \right)^{r/k}  \left( z \right)^{r(k-1)/k} 
\end{equation}

In the following we focus on the case $r=4$. In that case the four$-$point function $R_k(z)$ is given by (cf Appendix (\ref{4_points})) :
\begin{equation}
R_k(z) = 1 -\frac{4}{k} z + \frac{2(k-4)}{k^2}\left(\frac{4(k-1)^2}{(k-4)c} -1 \right) z^2  -\frac{4}{k} z^3 + z^4 \label{4 points}
\end{equation}

Injecting Eq.(\ref{4 points}) in Eq.(\ref{specialization}) yields immediately (up to a global normalization):

\begin{eqnarray}
a_{[\underbrace{4,4,\dots,4}_{k-1},4]} & = & 1 \\
a_{[\underbrace{4,4,\dots,4}_{k-1},3,1]} & = & -\frac{4}{k} \\
a_{[\underbrace{4,4,\dots,4}_{k-1},2,2]} & = &  \frac{2(k-4)}{k^2}\left( \frac{4(k-1)^2}{(k-4)c} -1 \right) 
\end{eqnarray}

Demanding that the expansion obeys the (HW) and (LW) properties further constraint the coefficients $a_{\lambda}$. As it turns out, this is sufficient to determine completely the polynomial $P_{2k}^{(k,4)}$ for $k \leq 5$. 

\subsubsection{k=3}
\label{App3}

There are %$\langle \left[ \begin{array}{c} 6+4 \\ 4 \end{array}\right]_q \mid q^{12} \rangle = 18$
$18$ monomials $m_{\lambda}$, and translation invariance (HW) yields $16$ \emph{independent} constraints, and we are left with a two dimensional space ((LW) is automatically satisfied in that subspace). Since $J_{[4,4,4]}^{-4/3}$ and $J_{[4,4,2,2]}^{-3}$ belong to this space, they form a basis :

>From the analysis above we know that :

\begin{eqnarray}
a_{[4,4,4]} & = & 1 \\
a_{[4,4,3,1]} & = & -\frac{4}{3} \\
a_{[4,4,2,2]} & = &  \frac{2}{9}\left( \frac{16}{c} + 1 \right) 
\end{eqnarray}

By comparing with the exact expansion of the Jacks  $J_{[4,4,4,0,0,0]}^{-4/3}$ and $J_{[4,4,2,2,0,0]}^{-3}$ one finally gets for a generic central charge $c$ :

\begin{eqnarray}
P_{6}^{(3,4)} (\{z_i\}) = &   J_{[4,4,4]}^{-4/3} + \frac{32}{9}\left(\frac{7}{40}+\frac{1}{c} \right) J_{[4,4,2,2]}^{-3} 
\end{eqnarray}

\subsubsection{k=5}
\label{App5}
In that case there are  %$\langle \left[ \begin{array}{c} 10+4 \\ 4 \end{array}\right]_q \mid q^{20} \rangle = 55$
$55$ monomials $m_{\lambda}$ and translation invariance yields $53$ \emph{independent} constraints, and we are again left with a two dimensional space.

But this time $J_{[4,4,4,4,4]}^{\alpha}$ is singular for $\alpha = -\frac{5+1}{4-1} = -2$ . This is possible since $r-1$ and $k+1$ must be coprime to ensure definiteness of the relevant Jacks. Nonetheless the polynomial $P_{10}^{(5,4)}$ is perfectly well defined by the parafermionic algebra of $\mathbb{Z}_{5}^{(2)}$.% and admits, up to a normalization, the following monomial expansion ( we have set $X = a_{[4,4,4,4,2,2]}  =   \frac{2}{25}\left( \frac{64}{c} -1 \right) $, and the partitions are labeled by lexical ordering ) : 

%\begin{eqnarray}
%a1 & = & 1, \\
%a2 & = & -4/5, \\
%a3 & = & X,\\
%a4 & = & -(1/2)*X+3/5,\\
%a5 & = & X-6/5,\\
%a6 & = & -(1/2)*X+3/5,\\
%a7 & = & (1/4)*X+13/10,\\
%a8 & = & -(1/4)*X-9/10,\\
%a9 & = & (3/4)*X-9/10,\\
%a10 & = & -5*X+6,\\
%a11 & = & 3*X+9/5, \\
%a12 & = & -(5/4)*X+3/2, \\
%a13 & = & (3/2)*X-9/5, \\
%a14 & = & 0, \\
%a15 & = & X-6/5, \\
%a16 & = & (3/4)*X-9/10, \\
%a17 & = & -4-2*X, \\
%a18 & = & -(5/4)*X+3/2, \\
%a19 & = & 16/5, \\
%a20 & = & 0, \\
%a21 & = & 40*X-48, \\
%a22 & = & -42/5+X, \\
%a23 & = & (1/2)*X-3/5, \\
%a24 & = & -10*X+12,\\ 
%a25 & = & 60-5*X,\\ 
%a26 & = & -(5/2)*X+3, \\
%a27 & = & 9*X-54/5, 
%\end{eqnarray}
%\begin{eqnarray}
%a28 & = & -5*X+6,\\ 
%a29 & = & (3/2)*X-9/5, \\
%a30 & = & 0, \\
%a31 & = & 32+16*X,\\
%a32 & = & (1/2)*X-3/5, \\
%a33 & = & -18-X,\\ 
%a34 & = & -40*X+48, \\
%a35 & = & -(5/2)*X+3,\\ 
%a36 & = & 171/5-(9/2)*X, \\
%a37 & = & 32*X-192/5,\\ 
%a38 & = & -147+(65/2)*X, \\
%a39 & = & -(105/2)*X+63,\\ 
%a40 & = & 1638-420*X,\\ 
%a41 & = & (315/2)*X-189, \\
%a42 & = & 0,\\ 
%a43 & = & 40*X-48, \\
%a44 & = & -10*X+12,\\ 
%a45 & = & -40*X+48,\\ 
%a46 & = & -1024,\\ 
%a47 & = & 9*X-54/5, \\
%a48 & = & 32*X-192/5,\\ 
%a49 & = & 1152/5+64*X,\\ 
%a50 & = & -(105/2)*X+63, \\
%a51 & = & -1008/5-120*X,\\ 
%a52 & = & (315/2)*X-189,\\ 
%a53 & = & 420+370*X,\\ 
%a54 & = & -2016-2100*X, \\
%a55 & = & 27216+28350*X
%\end{eqnarray}

This polynomial can be expressed using Jacks through :

\begin{equation}
P_{10}^{(5,4)} = \left( J_{[2,2,2,2,2]}^{-6} \right)^2 + \frac{128}{25} \left( \frac{1}{c} - \frac{7}{16}\right) \left(  J_{[4,4,4,4,2,2]}^{-5} -\frac{3}{8} J_{[4,4,4,3,2,2,1]}^{-5} + \frac{18}{7} J_{[4,4,4,2,2,2,2]}^{-5} + \frac{3}{11} J_{[4,4,3,3,2,2,1,1]}^{-5}  \right)
\end{equation}

\subsection{Computation of a particular $4-$point function for $\mathbb{Z}_{k}^{(4)}$}
\label{4_points}
In order to evaluate the $2k-$point correlators in $\mathbb{Z}_{k}^{(4)}$, it is convenient to know the exact form of the following $4-$point function $\langle \Psi_{-1}(\infty) \Psi_1(1) \Psi_1(z) \Psi_{-1}(0)\rangle$. It is straitforward to obtain it from the $\mathbb{Z}_k^{(r)}(c)$ OPEs :

\begin{eqnarray}
\Psi_1(z) \Psi_1(z') & = & \frac{\gamma}{\left(z-z'\right)^{2\frac{k-1}{k}}}\left( \Psi_2(z') + \left(z-z'\right)\frac{1}{2}  \partial \Psi_{2}(z') + \mathcal{O}((z-z')^2) \right) \\
 \Psi_1(z) \Psi_{-1}(z') & = &  \frac{1}{\left(z-z'\right)^{4\frac{k-1}{k}}}\left( 1 + \left(z-z'\right)^2\frac{4(k-1)}{kc}  T(z') + \mathcal{O}((z-z')^3) \right)
\end{eqnarray}

Since $\Psi_1(z)\sim z^{-2(k-1)/k} $ when $z \rightarrow \infty$, the following polynomial is of degree at most $4$ : 

\begin{eqnarray}
R_k(z) & = & \langle \Psi_{-1}(\infty) \Psi_1(1) \Psi_1(z) \Psi_{-1}(0)\rangle \left(1-z\right)^{2\frac{k-1}{k}} z^{4\frac{k-1}{k}} \\
& = & a_0 + a_1 z + a_2 z^2 + a_3 z^3 + a_4 z^4 
\end{eqnarray}

The conformal transformation $z\rightarrow \frac{1}{z}$ gives

\begin{equation}
R_k(z) = z^4 R_k(\frac{1}{z})
\end{equation}

i.e. $a_4=a_0$ and $a_3=a_1$, so there are only 3 parameters left. It turns out that the dominant terms in the OPEs are sufficient to determine all $a_i$'s :

\begin{eqnarray}
\Psi_1(z) \Psi_1(z') & = & \frac{\gamma}{\left(z-z'\right)^{2\frac{k-1}{k}}}\left( \Psi_2(z') + \mathcal{O}((z-z')) \right) \\
 \Psi_1(z) \Psi_{-1}(z') & = &  \frac{1}{\left(z-z'\right)^{4\frac{k-1}{k}}}\left( 1 + \mathcal{O}((z-z')^2) \right)
\end{eqnarray}

By comparing the OPEs with the polynomial as $z \rightarrow 0$ one finds $a_0 =1$, $a_1=-\frac{4}{k}$. And doing so as $z \rightarrow 1$ gives $R_k(1)=\sum_i a_i = \gamma^2$

\begin{eqnarray}
R_k(z) & = & \langle \Psi_{-1}(\infty) \Psi_1(1) \Psi_1(z) \Psi_{-1}(0)\rangle \left(1-z\right)^{2\frac{k-1}{k}} z^{4\frac{k-1}{k}}  \\
& = & 1 -\frac{4}{k} z + \left(\gamma^2 - 2\frac{k-4}{k}  \right) z^2  -\frac{4}{k} z^3 + z^4
\end{eqnarray}

Finally, the algebra constant $\gamma$ and the central charge $c$ are related by the associativity of the OPEs in $\mathbb{Z}_{k}^{(2)}$ :
\begin{equation}
\gamma^2  = \frac{8(k-1)}{k^2}  \left( \frac{k-1}{c} +\frac{k-4}{4} \right)
\end{equation}
 
\begin{eqnarray}
R_k(z) & = & \langle \Psi_{-1}(\infty) \Psi_1(1) \Psi_1(z) \Psi_{-1}(0)\rangle \left(1-z\right)^{2\frac{k-1}{k}} z^{4\frac{k-1}{k}}  \\
& = & 1 -\frac{4}{k} z + 2\frac{k-4}{k^2}\left(\frac{4(k-1)^2}{(k-4)c} -1 \right) z^2  -\frac{4}{k} z^3 + z^4
\end{eqnarray}

\subsection{Conformal properties through the stress-energy tensor} 
\label{AppConf}

In this appendix we show how conformal invariance for correlators translates into (LW) and (HW) conditions fro the corresponding polynomial.
Let us consider a generic holomorphic $n-$point correlator in any CFT : 
\begin{equation}
\langle \prod_{i=1}^n \Phi_i (z_i)\rangle
\end{equation}

where the $\Phi_i$ are primary fields with conformal dimension $\Delta_i$. The behavior of the stress-energy tensor at infinity, namely $T(z) \sim \frac{1}{z^4}$, together with the conformal Ward identity for primary fields $T(z) \Phi_{i} (z_i) = \frac{\Delta_i}{(z-z_i)^2}\Phi_{i}(z_i) + \frac{1}{z-z_i}\partial \Phi_{i}(z_i) + \dots $ ensures the following properties :

\begin{equation}
 \frac{1}{2\pi i} \oint z^p \langle T(z) \prod_i \Phi_i (z_i)\rangle = 0 \quad \textrm{for } p\leq 2
\end{equation}
 
This gives for $p=0,1,2$ :
\begin{eqnarray}
\sum_i \partial_i  \langle  \prod_i \Phi_i (z_i)\rangle  & = & 0 \\
\sum_i \left( z_i \partial_i  + \Delta_i \right) \langle  \prod_i \Phi_i (z_i)\rangle  & = & 0 \\
\sum_i \left( z_i^2 \partial_i  + 2 z_i \Delta_i \right) \langle  \prod_i \Phi_i (z_i)\rangle  & = & 0 
\end{eqnarray}

We now consider the special case of a parafermionic CFT $\mathbb{Z}^{(r)}_k$, where the parafermionic fields $\Psi_q$ have dimension $\Delta_q=\frac{r}{2}\frac{q(k-q)}{k}$. We are interested in the properties of the following polynomial : 

\begin{eqnarray}
P_n^{(k,r)} (\{z_i\})  & \hat{=} & \langle \Psi(z_1) \hdots  \Psi(z_{n}) \rangle \prod_{i<j} \left( z_i-z_j \right)^{2\Delta_1-\Delta_2} \nonumber \\
& = & \langle  \Psi(z_1) \hdots  \Psi(z_{n}) \rangle \prod_{i<j} \left(z_i-z_j \right)^{r/k}
\end{eqnarray}

 Introducing the standard notation $\hat{E}_p = \sum_{i=1}^{n} z_i^p \partial_i $ we get : 

\begin{eqnarray}
\hat{E}_0 \left( \langle  \prod_i \Psi (z_i)\rangle \right) & = & 0 \\
\hat{E}_1 \left( \langle  \prod_i \Psi (z_i)\rangle \right) & = & - n \Delta_1 \langle  \prod_i \Psi (z_i)\rangle \\
\hat{E}_2 \left( \langle  \prod_i \Psi (z_i)\rangle \right) & = & - 2\Delta_1 \left( \sum_i z_i\right) \langle  \prod_i \Psi (z_i)\rangle  
\end{eqnarray}

The phase term $\prod_{i<j} z_{ij}^{\frac{r}{k}}$ also behaves nicely:
\begin{eqnarray}
\hat{E}_0  \left( \prod_{i<j} z_{ij}^{r/k} \right)  & = & 0 \\
\hat{E}_1  \left( \prod_{i<j} z_{ij}^{r/k} \right)  & = & \frac{r}{k} \frac{n(n-1)}{2}  \prod_{i<j} z_{ij}^{\gamma} \\
\hat{E}_2  \left( \prod_{i<j} z_{ij}^{r/k} \right)  & = & \frac{r}{k} (n-1)\left( \sum_i z_i \right) \prod_{i<j} z_{ij}^{\gamma}  
\end{eqnarray}

This gives for the polynomials :

\begin{eqnarray}
\hat{E}_0  \left( P_n^{(k)} \right)  & = & 0 \\
\hat{E}_1  \left( P_n^{(k)} \right)  & = & \frac{1}{2}n N_{\Phi} P_n^{(k)} \\
\hat{E}_2  \left( P_n^{(k)} \right)  & = & N_{\Phi} \left( \sum_i z_i \right)P_n^{(k)} 
\end{eqnarray}

i.e.  $N_{\Phi}=r(\frac{n}{k}-1)$ is the degree in $z_i$ and $\frac{1}{2}n N_{\Phi}=\frac{r}{2}n(\frac{n}{k}-1)$ is the total degree of the polynomial.

These properties are nothing but the Highest Weight (HW), homogeneity and Lowest weight (LW) conditions. They are inherited from the underlying CFT, and are just a rewriting of the Ward identities associated to the global conformal transformations $SL(2,\mathbb{C})$.

\subsection{Equivalently through conformal transformations}

All these properties can be integrated by applying the finite form of these  $SL(2,\mathbb{C})$ transformations.
%Under $(z,\bar{z}) \rightarrow \left(f(z),\overline{f(z)} \right)$ primary fields transform like :
%\begin{equation}
%\Phi_{\Delta,\overline{\Delta}}(z,\bar{z}) \rightarrow \left( \frac{df(z)}{dz} \right)^{\Delta}\left( \frac{\overline{df(z)}}{dz}  \right)^{\overline{\Delta}} \Phi_{\Delta,\overline{\Delta}}(f(z),\overline{f(z)})
%\end{equation}

Under $z \rightarrow f(z)$ primary fields transform like :
\begin{equation} \Phi_{\Delta}(z) \rightarrow \left( \frac{df(z)}{dz} \right)^{\Delta} \Phi_{\Delta}(f(z))\end{equation}

Translations, dilatations and special conformal transformation yield :
\begin{itemize}
\item $z \rightarrow z-z_{n}$
\begin{equation}
P_n^{(k)}\left(z_1, \dots z_{n}\right) = P_N^{(k)} \left(z_1- z_{n}, z_2-z_{n} , \dots ,z_{n-1}-z_{n},0 \right)
\end{equation}
i.e. $P$ is translation invariant (HW)

\item $z \rightarrow \lambda z$ 
\begin{equation}
 P_n^{(k)} \left(\lambda z_1,\lambda z_2 , \dots ,\lambda z_{n} \right) = \lambda^{\frac{1}{2} n N_{\Phi}}   P_n^{(k)}\left(z_1, \dots z_{n}\right)
\end{equation}
i.e. $P$ is homogeneous with total degree $\frac{1}{2} n N_{\Phi}$

\item $z \rightarrow -\frac{1}{z}$ 
\begin{equation}
P_n^{(k)}\left(z_1, \dots z_{n}\right) = \prod_{i=1}^{n} \left(z_{i} \right)^{N_{\Phi}} P_n^{(k)} \left( -\frac{1}{z_{1}} , \dots ,-\frac{1}{z_{n}} \right) 
\end{equation}
which is just the integrated version of the (LW) condition

%Proof : if $P_n^{(k)}\left(z_1, \dots z_{n}\right) = \prod_{i=1}^{n} \left(z_{i} \right)^{N_{\Phi}} P_n^{(k)} \left( -\frac{1}{z_{1}} , \dots ,-\frac{1}{z_{n}} \right)$ then  \begin{eqnarray} \sum_i z_i^2 \partial_i P(z_i) & = &  \sum_i z_i^2 \partial_i \left( \prod_{i=1}^{n} \left(z_{i} \right)^{N_{\Phi}}  P_n^{(k)} \left(-\frac{1}{z_i}\right) \right) \\ & = &  \sum_i z_i^2 \partial_i \left( \prod_{i=1}^{n} \left(z_{i} \right)^{N_{\Phi}} \right) P_n^{(k)} \left(-\frac{1}{z_i}\right) +  \prod_{i=1}^{n} \left(z_{i} \right)^{N_{\Phi}} \sum_i z_i^2 \partial_i \left( P_n^{(k)} \left(-\frac{1}{z_i}\right) \right) \\ & = & N_{\Phi} \left( \sum_i z_i \right) P_n^{(k)}(z_i) +  \prod_{i=1}^{n} \left(z_{i} \right)^{N_{\Phi}} \left( \underbrace{\sum_i  \partial_i P_n^{(k)}}_{0} \right) \left(-\frac{1}{z_i}\right) \\ & = & N_{\Phi} \left( \sum_i z_i \right) P_n^{(k)}(z_i) \end{eqnarray} 
   %    if $\hat{E}_2 \left( P_n^{(k)} \right) = N_{\Phi} \left( \sum_i z_i \right) P_n^{(k)}$ then

Assuming $r$ even, the total degree $\frac{1}{2}n N_{\Phi}$ is even, and $P(z_i)=P(-z_i)$. Expanding the polynomial in the monomial basis, the last property boils down to $a_{\lambda} = a_{\lambda^t}$ where $\lambda^t$ is the young tableau complementary to $\lambda$ : 
\begin{equation}
 \lambda^t_i=N_{\Phi} - \lambda_{n-i+1} 
\end{equation}

For instance $[4,4,1,1,1,1]^t=[3,3,3,3]$
\end{itemize}

\end{document}